\documentclass[fleqn,usenatbib]{mnras}

\usepackage{newtxtext,newtxmath}
\usepackage{lineno}

\usepackage[T1]{fontenc}
\usepackage{ae,aecompl}
\usepackage{multirow}

\usepackage{graphicx}	
\usepackage{amsmath}	
\usepackage{upgreek}

\usepackage{ulem}
\usepackage{url}
\usepackage{xspace}
\usepackage{comment}
\usepackage[flushleft]{threeparttable}
\graphicspath{ {./figures/} }

\newcommand{\ms}{\mbox{m\,s$^{-1}$}\xspace}
\newcommand{\kms}{\mbox{km\,s$^{-1}$}\xspace}

\newcommand{\rhoe}{\mbox{$\rho_{\oplus}$}\xspace}

\newcommand{\masyr}{\mbox{mas\,yr$^{-1}$}\xspace}
\newcommand{\msun}{$M_{\odot}$\xspace}

\newcommand{\rsun}{$R_{\odot}$\xspace}

\newcommand{\mearth}{$M_\oplus$\xspace}
\newcommand{\rearth}{$R_\oplus$\xspace}

\newcommand{\Rmnum}[1]{\expandafter\@slowromancap\romannumeral #1@}

\newcommand{\mstar}{\ensuremath{M_{\star}}\xspace}
\newcommand{\rstar}{\ensuremath{R_{\star}}\xspace}

\newcommand{\m}{$[$m/H$]$\xspace}
\newcommand{\teff}{\ensuremath{T_{\mathrm{eff}}}\xspace}  
\newcommand{\logg}{\ensuremath{\log g}\xspace} 
\newcommand{\vsini}{\ensuremath{v \sin i}\xspace}

\newcommand{\Mp}{\ensuremath{M_{p}}\xspace} 

\newcommand{\Rp}{$R_{p}$\xspace}

\newcommand{\RpRs}{$R_p/R_\star$\xspace}
\newcommand{\aRs}{$a/R_\star$\xspace}
\newcommand{\To}{$T_0$\xspace}
\newcommand{\imppar}{$b$\xspace}
\newcommand{\Teq}{$T_{\mathrm{eq}}$\xspace}
\newcommand{\Me}{\ensuremath{M_{\oplus}}\xspace}

\newcommand{\Krv}{$K_\mathrm{RV}$\xspace}
\newcommand{\ecos}{$\sqrt{e}\,$cos\,$\omega$\xspace}
\newcommand{\esin}{$\sqrt{e}\,$sin\,$\omega$\xspace}
\newcommand{\Tdur}{$T_{14}$\xspace}
\newcommand{\rhop}{$\rho_{p}$\xspace}

\newcommand{\target}{HIP~113103\xspace}
\newcommand{\targetb}{HIP~113103~b\xspace}
\newcommand{\targetc}{HIP~113103~c\xspace}
\newcommand{\ticid}{TIC~121490076\xspace}

\newcommand{\rprsb}{$0.0242_{-0.0008}^{+0.0013}$\xspace}
\newcommand{\tcb}{$1325.5966_{-0.0024}^{+0.0033}$\xspace}
\newcommand{\periodb}{$7.610303 \pm 0.000018$\xspace}
\newcommand{\incb}{$ 88.23_{-0.14}^{+0.18}$\xspace}
\newcommand{\esinomegab}{$-0.12_{-0.32}^{+0.31}$\xspace}
\newcommand{\ecosomegab}{$0.18_{-0.45}^{+0.51}$\xspace}

\newcommand{\rpb}{$1.829_{-0.067}^{+0.096}$\xspace}
\newcommand{\ab}{$0.06899_{-0.00023}^{+0.00029}$\xspace}
\newcommand{\abvsrb}{$21.39_{-0.13}^{+0.10}$\xspace}
\newcommand{\bb}{$0.656_{-0.084}^{+0.070}$\xspace}
\newcommand{\eccentricityb}{$0.17_{-0.13}^{+0.17}$\xspace}
\newcommand{\omegab}{$-10_{-140}^{+120}$\xspace} 
\newcommand{\teqb}{$721\pm10$\xspace} 
\newcommand{\tfourteenb}{$0.0891_{-0.0068}^{+0.0075}$\xspace}
\newcommand{\kvb}{$2.34 \pm 0.73^{*}$\xspace}
\newcommand{\rprsbsq}{$0.000596_{-0.000051}^{+0.000062}$\xspace}

\newcommand{\rhob}{$0.96^{+0.15}_{-0.22}$\xspace}

\newcommand{\rprsc}{$0.0303_{-0.0010}^{+0.0014}$\xspace}
\newcommand{\tcc}{$1337.0559 \pm 0.0019$\xspace}
\newcommand{\periodc}{$14.245648 \pm 0.000019$\xspace}
\newcommand{\incc}{$89.24_{-0.22}^{+0.40}$\xspace}
\newcommand{\esinomegac}{$0.21_{-0.18}^{+0.13}$\xspace}
\newcommand{\ecosomegac}{$-0.31_{-0.25}^{+0.23}$\xspace}

\newcommand{\rpc}{$2.40_{-0.08}^{+0.10}$\xspace}
\newcommand{\ac}{$0.10479_{-0.00035}^{+0.00045}$\xspace}
\newcommand{\acvsrc}{$32.49_{-0.19}^{+0.15}$\xspace}
\newcommand{\bc}{$0.614_{-0.063}^{+0.028}$\xspace}
\newcommand{\eccentricityc}{$0.17_{-0.13}^{+0.17}$\xspace}
\newcommand{\omegac}{$-70_{-60}^{+100}$\xspace}
\newcommand{\teqc}{$585\pm10$\xspace}
\newcommand{\tfourteenc}{$0.1764_{-0.0050}^{+0.0091}$\xspace}
\newcommand{\kvc}{$2.67 \pm 0.58^{*}$\xspace}
\newcommand{\rprscsq}{$0.001051_{-0.000087}^{+0.00011}$\xspace}
\newcommand{\rhoc}{$0.60^{+0.054}_{-0.091}$\xspace}

\newcommand{\ra}{22:54:17.37\xspace}
\newcommand{\dec}{$-$43:00:37.25\xspace}
\newcommand{\parallax}{$ 21.61785\pm 0.00024$\xspace}
\newcommand{\gaiapmra}{$1.995 \pm 0.020$\xspace}
\newcommand{\gaiapmdec}{$27.384 \pm 0.021$\xspace}
\newcommand{\hippara}{$2.2 \pm 1.5$\xspace}
\newcommand{\hippadec}{$27.1 \pm 1.3$\xspace}
\newcommand{\gavgpmra}{$2.032 \pm 0.048$\xspace}
\newcommand{\gavgpmdec}{$27.396 \pm 0.038$\xspace}
\newcommand{\tessmag}{$8.9988 \pm 0.0063$\xspace}

\newcommand{\bmag}{$10.907 \pm 0.033$\xspace}
\newcommand{\vmag}{$9.95 \pm 0.03$\xspace}
\newcommand{\jmag}{$8.195 \pm 0.03$\xspace}
\newcommand{\hmag}{$ 7.67\pm 0.042$\xspace}
\newcommand{\kmag}{$7.557 \pm 0.031$\xspace}
\newcommand{\gmag}{$9.6175 \pm 0.0018$\xspace}
\newcommand{\gmagbp}{$10.1491 \pm 0.0033$\xspace}
\newcommand{\gmagrp}{$8.9353 \pm 0.0039$\xspace}
\newcommand{\wonemag}{$7.398 \pm 0.033$\xspace}
\newcommand{\wtwomag}{$7.538 \pm 0.02$\xspace}
\newcommand{\wthreemag}{$7.489 \pm 0.017$\xspace}
\newcommand{\wfourmag}{$7.395 \pm 0.132$\xspace}

\newcommand{\mhstar}{$0.761 \pm 0.038$\xspace}
\newcommand{\rhstar}{$0.742 \pm 0.013$\xspace}
\newcommand{\teffstar}{$4930\pm100$\xspace}
\newcommand{\loggstar}{$4.6 \pm 0.1$\xspace}
\newcommand{\metalicity}{$-0.1 \pm 0.1$\xspace}
\newcommand{\vsinistar}{$3 \pm 1$\xspace}

\newcommand{\age}{$470_{-110}^{+170}$\xspace}

\newcommand{\dstar}{$46.212 \pm 0.086$\xspace}
\newcommand{\ukms}{$5.00 \pm 0.17$\xspace}
\newcommand{\vkms}{$4.635 \pm 0.031$\xspace}
\newcommand{\wkms}{$-13.89 \pm 0.32$\xspace}

\newcommand{\tess}{\textit{TESS}\xspace}
\newcommand{\cheops}{\textit{CHEOPS}\xspace}

\newcommand{\gaia}{\textit{Gaia}\xspace}
\newcommand{\wise}{\textit{WISE}\xspace}
\newcommand{\hipp}{\textit{Hipparcos}\xspace}
\newcommand{\twinkle}{\textit{Twinkle}\xspace}
\newcommand{\jwst}{\textit{JWST}\xspace}

\newcommand{\hst}{\textit{HST}\xspace}

\usepackage{scalerel}
\usepackage{tikz}
\usetikzlibrary{svg.path}

\definecolor{orcidlogocol}{HTML}{A6CE39}
\tikzset{
  orcidlogo/.pic={
    \fill[orcidlogocol] svg{M256,128c0,70.7-57.3,128-128,128C57.3,256,0,198.7,0,128C0,57.3,57.3,0,128,0C198.7,0,256,57.3,256,128z};
    \fill[white] svg{M86.3,186.2H70.9V79.1h15.4v48.4V186.2z}
                 svg{M108.9,79.1h41.6c39.6,0,57,28.3,57,53.6c0,27.5-21.5,53.6-56.8,53.6h-41.8V79.1z M124.3,172.4h24.5c34.9,0,42.9-26.5,42.9-39.7c0-21.5-13.7-39.7-43.7-39.7h-23.7V172.4z}
                 svg{M88.7,56.8c0,5.5-4.5,10.1-10.1,10.1c-5.6,0-10.1-4.6-10.1-10.1c0-5.6,4.5-10.1,10.1-10.1C84.2,46.7,88.7,51.3,88.7,56.8z};
  }
}

\newcommand\orcid[1]{\href{https://orcid.org/#1}{\mbox{\scalerel*{
\begin{tikzpicture}[yscale=-1,transform shape]
\pic{orcidlogo};
\end{tikzpicture}
}{|}}}}

\title[Discovery of two mini-Neptunes Around \target{}]{Two mini-Neptunes Transiting the Adolescent K-star \target{} Confirmed with \tess{} and \cheops{}}

\author[N. Lowson]{\parbox{\textwidth}
{N. Lowson\orcid{0000-0001-6508-5736},$^{1}$\thanks{E-mail: \texttt{nataliea.lowson@usq.edu.au}}
G. Zhou\orcid{0000-0002-4891-3517},$^{1}$
C. X. Huang\orcid{0000-0003-0918-7484},$^{1}$
D. J. Wright\orcid{0000-0001-7294-5386},$^{1}$
B. Edwards\orcid{0000-0002-5494-3237},$^{2}$
E. Nabbie\orcid{0000-0003-0571-2245},$^{1}$
A. Venner\orcid{0000-0002-8400-1646},$^{1}$
S. N. Quinn\orcid{0000-0002-8964-8377},$^{3}$
K. A. Collins\orcid{0000-0001-6588-9574},$^{3}$
E. Gillen\orcid{0000-0003-2851-3070},$^{4,5,6}$
M. Battley\orcid{0000-0002-1357-9774},$^{7}$
A. Triaud\orcid{0000-0002-5510-8751},$^{8}$
C. Hellier\orcid{0000-0002-3439-1439},$^{9}$
S. Seager\orcid{0000-0002-6892-6948},$^{10, 11, 12}$
J. N. Winn\orcid{0000-0002-4265-047X},$^{13}$
J. M. Jenkins\orcid{0000-0002-4715-9460},$^{14}$
B. Wohler\orcid{0000-0002-5402-9613},$^{14, 15}$
A. Shporer\orcid{0000-0002-1836-3120},$^{10}$
R. P. Schwarz\orcid{0000-0001-8227-1020},$^{3}$
F. Murgas\orcid{0000-0001-9087-1245},$^{16, 17}$
E. Pallé\orcid{0000-0003-0987-1593},$^{16, 17}$
D. R. Anderson\orcid{0000-0001-7416-7522},$^{18, 19}$
R. G. West\orcid{0000-0001-6604-5533},$^{18, 19}$
R. A. Wittenmyer\orcid{0000-0001-9957-9304},$^{1}$
B. P. Bowler\orcid{0000-0003-2649-2288},$^{20}$ 
J. Horner\orcid{0000-0002-1160-7970},$^{1}$
S. R. Kane\orcid{0000-0002-7084-0529},$^{21}$
J. Kielkopf\orcid{0000-0003-0497-2651},$^{22}$
P. Plavchan\orcid{0000-0002-8864-1667},$^{23}$ 
H. Zhang\orcid{0000-0003-3491-6394},$^{24}$
T. Fairnington\orcid{0000-0002-0692-7822},$^{1}$
J. Okumura,\orcid{0000-0002-4876-8540}$^{1}$
M. W. Mengel\orcid{0000-0002-7830-6822},$^{1}$
B. C. Addison\orcid{0000-0003-3216-0626}$^{1}$
}
\\
\\
$^{1}$Centre for Astrophysics, University of Southern Queensland, 499-565 West Street, Toowoomba, QLD 4350, Australia \\
$^{2}$SRON, Netherlands Institute for Space Research, Niels Bohrweg 4, NL-2333 CA, Leiden, The Netherlands \\
$^{3}$Center for Astrophysics \textbar \ Harvard \& Smithsonian, 60 Garden Street, Cambridge, MA 02138, USA \\
$^{4}$Winton Fellow \\
$^{5}$Astronomy Unit, Queen Mary University of London, Mile End Road, London E1 4NS, UK \\
$^{6}$Astrophysics Group, Cavendish Laboratory, J.J. Thomson Avenue, Cambridge CB3 0HE, UK \\
$^{7}$Observatoire astronomique de l’Université de Genève, Chemin Pegasi 51, 1290 Versoix, Switzerland \\
$^{8}$School of Physics \& Astronomy, University of Birmingham, Edgbaston, Birmingham, B15 2TT, UK \\
$^{9}$Astrophysics Group, Keele University, Keele ST5 5BG, UK \\
$^{10}$Department of Physics and Kavli Institute for Astrophysics and Space Research, Massachusetts Institute of Technology, Cambridge, MA 02139, USA \\
$^{11}$Department of Earth, Atmospheric and Planetary Sciences, Massachusetts Institute of Technology, 77 Massachusetts Ave, Cambridge, MA 02139, USA \\
$^{12}$Department of Aeronautics and Astronautics, Massachusetts Institute of Technology, 77 Massachusetts Avenue, Cambridge, MA 02139, USA \\
$^{13}$Department of Astrophysical Sciences, Princeton University, 4 Ivy Lane, Princeton, NJ 08544, USA \\
$^{14}$NASA Ames Research Center, Moffett Field, CA 94035, USA \\
$^{15}$SETI Institute, Mountain View, CA 94043, USA \\
$^{16}$Instituto de Astrof\'isica de Canarias (IAC), E-38205 La Laguna, Tenerife, Spain \\
$^{17}$Departamento de Astrof\'isica, Universidad de La Laguna (ULL), E-38206 La Laguna, Tenerife, Spain \\
$^{18}$Department of Physics, University of Warwick, Gibbet Hill Road, Coventry CV4 7AL, UK \\
$^{19}$Centre for Exoplanets and Habitability, University of Warwick, Gibbet Hill Road, Coventry CV4 7AL, UK \\
$^{20}$Department of Astronomy, The University of Texas at Austin, TX 78712, USA \\
$^{21}$Department of Earth and Planetary Sciences, University of California, Riverside, CA 92521, USA \\
$^{22}$Department of Physics and Astronomy, University of Louisville, Louisville, KY 40292, USA \\
$^{23}$George Mason University, 4400 University Drive MS 3F3, Fairfax, VA 22030, USA \\
$^{24}$Shanghai Astronomical Observatory, Chinese Academy of Sciences, Shanghai 200030, China \\
}

\date{Accepted XXX. Received YYY; in original form ZZZ}

\pubyear{2022}

\begin{document}
\label{firstpage}
\pagerange{\pageref{firstpage}--\pageref{lastpage}}
\maketitle

\clearpage

\begin{abstract}
We report the discovery of two mini-Neptunes in near 2:1 resonance orbits (P = 7.610303 d for \targetb{} and P = 14.245651 d for \targetc{}) around the adolescent K-star \target{} (\ticid{}). The planet system was first identified from  the \tess{} mission, and was confirmed via additional photometric and spectroscopic observations, including a $\sim$17.5 hour observation for the transits of both planets using ESA \cheops{}. We place $\leq4.5$ min and $\leq2.5$ min limits on the absence of transit timing variations over the three year photometric baseline, allowing further constraints on the orbital eccentricities of the system beyond that available from the photometric transit duration alone. With a planetary radius of~\Rp=\,\rpb \rearth, \targetb{} resides within the radius gap, and this might provide invaluable information on the formation disparities between super-Earths and mini-Neptunes. Given the larger radius~\Rp= \rpc \rearth for \targetc{}, and close proximity of both planets to \target{}, it is likely that \targetb{} might have lost (or is still losing) its primordial atmosphere. We therefore present simulated atmospheric transmission spectra of both planets using \jwst{}, \hst{}, and \twinkle{}. It demonstrates a potential metallicity difference (due to differences in their evolution) would be a challenge to detect if the atmospheres are in chemical equilibrium. As one of the brightest multi sub-Neptune planet systems suitable for atmosphere follow up, \targetb{} and \targetc{} could provide insight on planetary evolution for the sub-Neptune K-star population.
\end{abstract}

\begin{keywords}
techniques: spectroscopic – techniques: radial velocities – planets and satellites: detection – stars: individual:
TIC121490076
\end{keywords}


\section{Introduction}
\label{sec:intro}

Super-Earths (1 \rearth< \Rp $\leq$ 1.5 \rearth) and mini-Neptunes (1.5 \rearth< \Rp $\leq$ 4 \rearth) are the most common planets found around sun-like stars (referred to as sub-Neptunes hereafter), especially those residing in close-in orbits~\citep{2012_Howard, 2013_Fressin, 2022_Bergsten}, despite having no analogues in our own Solar System. These planets bridge the gap between rocky Earth-like worlds and gaseous Neptunes~\citep[e.g.][]{2017_Fulton}. The \textit{Transiting Exoplanet Survey Satellite}~\citep[\tess{}, ][]{2015JATIS...1a4003R} mission continues to expand our repertoire for sub-Neptunes, in particular those orbiting bright nearby stars. These discoveries have led to precise radius and mass constraints for a significant number of sub-Neptunes~\citep[e.g.][]{2019_Dragomir, 2019_Gandolfi, 2020_Cloutier, 2021_MacDougall, 2021_Sozzetti, 2022_Gan, 2022_Lubin}, as well as the possibility of in-depth atmospheric characterisations that reveal the origins and evolutionary pathways of this population~\citep[e.g.][]{2021_Osborn, 2022_Kawauchi}.

Hypothesised planet formation pathways for sub-Neptunes (see \cite{{2021_Bean}} for more information) will exhibit observable differences that are accessible with the new generation of space and ground based facilities~\citep[e.g.][]{2016_Greene, 2018_Tinetti}. Depending on what occurs after dissipation of the gas disk, sub-Neptunes may not contain enough mass to gravitationally maintain their primordial atmosphere~\citep[e.g.][]{1986_Walker, 2012_Lopez, 2018_Ginzburg, 2020_Kite_a, 2020_Kite_b}. The rate of mass-loss post-formation is strongly dependent on the irradiation the planets receive from their host stars. Planets receiving strong XUV irradiation may be more likely to lose their primordial envelope~\citep[e.g.][]{2012_Owen, 2015_Howe, 2020_Mordasini, 2022_Ketzer}.

The next generation space-based telescopes \citep[commencing with the \textit{James Webb Space Telescope}, \jwst{},][]{2016_Greene} will be capable of characterising the atmospheres of sub-Neptunes, and many are prioritising wavelength regions towards the infrared~\citep[e.g.][]{2018_Tinetti, 2022_twinkle}. Obstruction by haze and clouds are minimised at longer wavelengths, and early \jwst{} observations have already demonstrated its invaluable retrieval capabilities for exoplanets atmospheres~\citep[e.g.][]{2023_Ahrer, 2023_Alderson, 2023_Feinstein, 2023_Rustamkulov, 2023_Tsai}. Prior to the launch of these next generation telescopes, some attempts of measuring sub-Neptune atmospheres have resulted in observations obscured by haze~\citep[e.g.][]{2014_Kreidberg, 2021_Mugnai}, however there have been notable exceptions which suggest predominant H/He envelopes~\citep[e.g.][]{2019_Benneke, 2019_Tsiaras, 2022_Orell-Miquel, edwards_pop}. Due to their size, observing sub-Neptune atmospheres is challenging in comparison to their larger Jovian counterparts, particularly around FGK stars. Therefore, the most suitable population of sub-Neptunes for atmosphere analysis are those residing in close orbits to bright host stars. In the known FGK planet population, there are only a handful of sub-Neptunes that meet these requirements~\citep[e.g.][]{2011_Winn, 2018_Gandolfi, 2019_Dragomir, 2020_Teske}, with samples dwindling further when only considering multi sub-Neptune planet systems~\citep[e.g.][]{2018_Rodriguez, 2021_Delrez, 2021_Scarsdale, 2022_Barragan}. It is therefore vital to identify sub-Neptunes with short periods around bright stars, as these candidates will lead the research towards understanding the formation pathways of this vast sub-class.

In this paper, we report the discovery of two sub-Neptunes that orbit at a 2:1 resonance around the bright K-star \target{} (\ticid). The initial observations with \tess{} and subsequent follow up with the \textit{CHaracterising ExOPlanets Satellite}~\citep[\cheops{},][]{2021_cheops} and ground-based facilities are outlined in Section~\ref{sec:obs}, while our global model fit to constrain the physical parameters of each planet are outlined in Section~\ref{sec:global_model}. The physical properties of \target{} are discussed in Section~\ref{sec:stellar}, while the elimination of false positive scenarios are outlined in Section~\ref{sec:FP}. Our Results and Discussion are presented in Section~\ref{sec:discussion} followed by our Conclusion in Section~\ref{sec:conclusion} respectively.

\section{Observations}
\label{sec:obs}

\subsection{Candidate identification with \tess{}} 
\label{sec:tess}

The transiting planets around \target{} were first identified by observations from \tess{}. Observations for \target{} were obtained via the 30 minute cadence Full Frame Images (FFI) from Sector 1 Camera 2, and via 10 minute FFIs and 2 minute target pixel stamps from Sector 28 Camera 2. 

The transit signals around \target{} were identified as part of a search for planets around young active field stars \citep{2021AJ....161....2Z} via public FFI light curves from the MIT Quick look pipeline~\citep{2020RNAAS...4..204H,2020RNAAS...4..206H}. The target star was identified as a potential young star via its high amplitude rotational modulation using the 10 minute FFI light curves from Sector 28. The combined FFI light curves of sector 1 and 28 were first modeled and detrended via a spline interpolation \citep{2014PASP..126..948V}, and searched for transit signals via the box-least-squared (BLS) procedure \citep{2002A&A...391..369K}. Two candidate signals are detected by BLS, one at $\approx$ 7.61 day with a signal to noise of 14, the other at $\approx$ 14.24 day with a signal to noise of 12.79. Both signals crossed the recommended threshold to be classified as a threshold crossing event (TCE) as defined by the TOI team \citep{2021ApJS..254...39G}. We vetted the data for both TCEs to rule out astrophysical false positives due to blending from nearby eclipsing binaries outside of the center pixel. We found that transit depth derived from different apertures are similar, and found no obvious blending sources when examining light curves from individual pixels in and around the aperture. We then promoted both TCEs for further follow up via \cheops{} (Section~\ref{sec:cheops}) and ground based instruments (Section~\ref{sec:ground_FU} and Section~\ref{sec:spec}).

\begin{figure*}
    \centering
    \includegraphics[width=14cm]{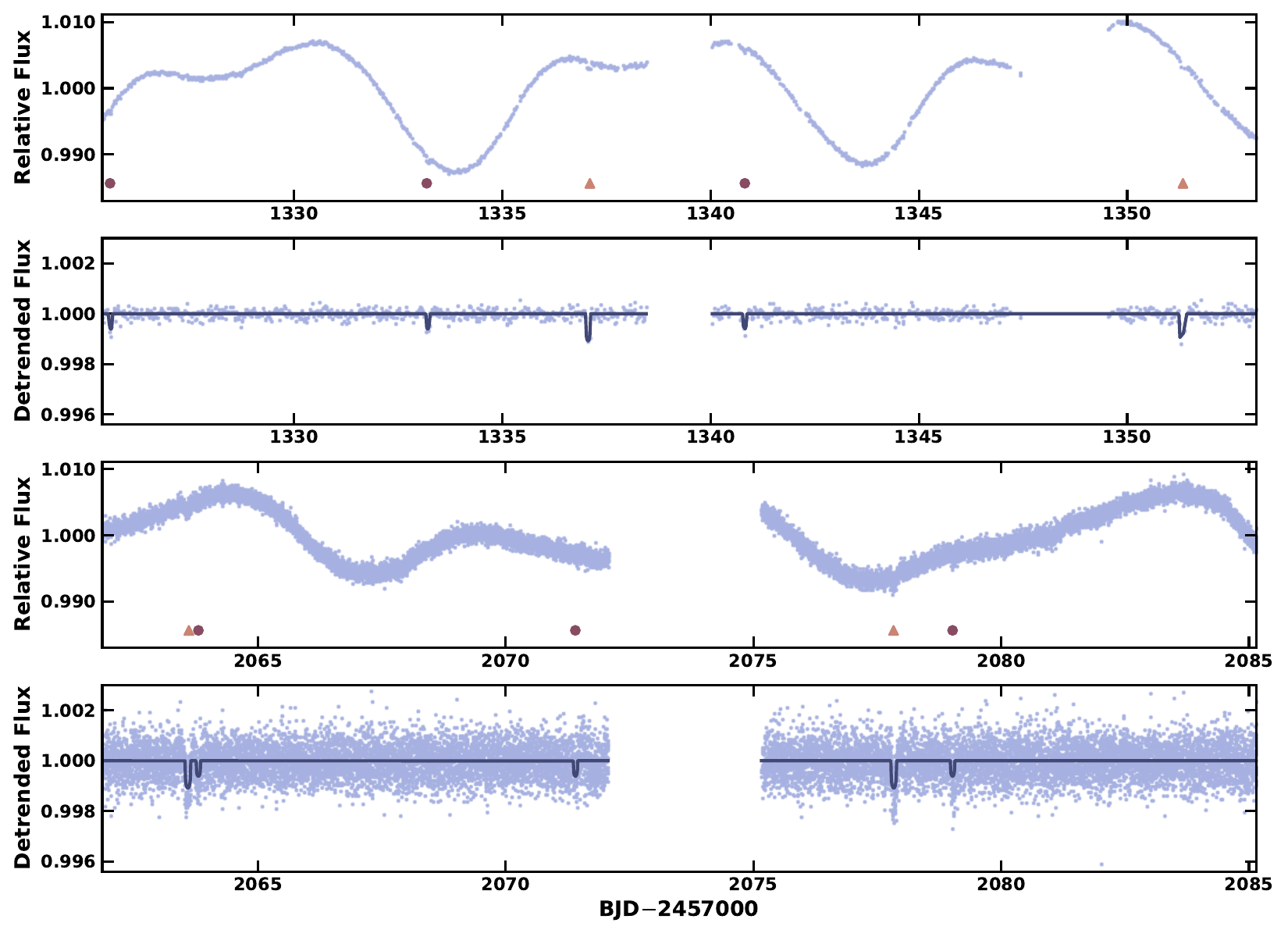}
    \caption{The \tess{} light curves before and after the removal of spot modulated rotational signals. The light curves of \target{} from Sector 1 via FFI observations were observed at 30 minute cadence (Panel 1 and 2), and from Sector 28 TPF observations at 2 minute (Panel 3 and 4).  Transits by \targetb{} and \targetc{} are illustrated via a circle and triangle respectively. The best fit transit model is displayed in navy.}
    \label{fig:lc_tess}
\end{figure*}

To refine the orbital and physical characteristics of the planets in our global model (Section~\ref{sec:global_model}), we use of the debelended Sector 28 target pixel stamp (TPF) two minute cadence Simple Aperture Photometry (SAP) light curves \citep{twicken:PA2010SPIE,morris2020}, performing the deblending using the contamination keywords in the TPF files. These light curves originate  from the Science Processing Operations Center \citep[SPOC,][]{2016SPIE.9913E..3EJ} at NASA Ames Research Center, and are made available via the the Mikulski Archive for Space Telescopes (MAST)\footnote{\href{https://mast.stsci.edu/portal/Mashup/Clients/Mast/Portal.html}{https://archive.stsci.edu/}}. 
\target{} exhibits significant rotational modulation due to spot activity on the stellar surface. To ensure proper propagation of the uncertainties associated with these noise sources, we model the rotational modulation and spacecraft systematics alongside the transiting planet signals. 
We use the deblended simple aperture SPOC light curves in this simultaneously detrending procedure.
Following \citet{2019ApJ...881L..19V}, we describe these signals as a linear combination of the spacecraft quarternions, the top seven covariant basis vectors, and a set of 20 cosine and sine functions at frequencies up to the \tess orbital period of 13 days~\citep[also see][]{2010_Mazeh, 2013_Huang}. Figure~\ref{fig:lc_tess} shows the \tess{} discovery light curve before and after the removal of the stellar and instrumental effects. Figure~\ref{fig:phasefold} shows the phase folded \textit{TESS} transit light curves for each planet.

\begin{figure*}
    \centering
    \includegraphics[width=\textwidth]{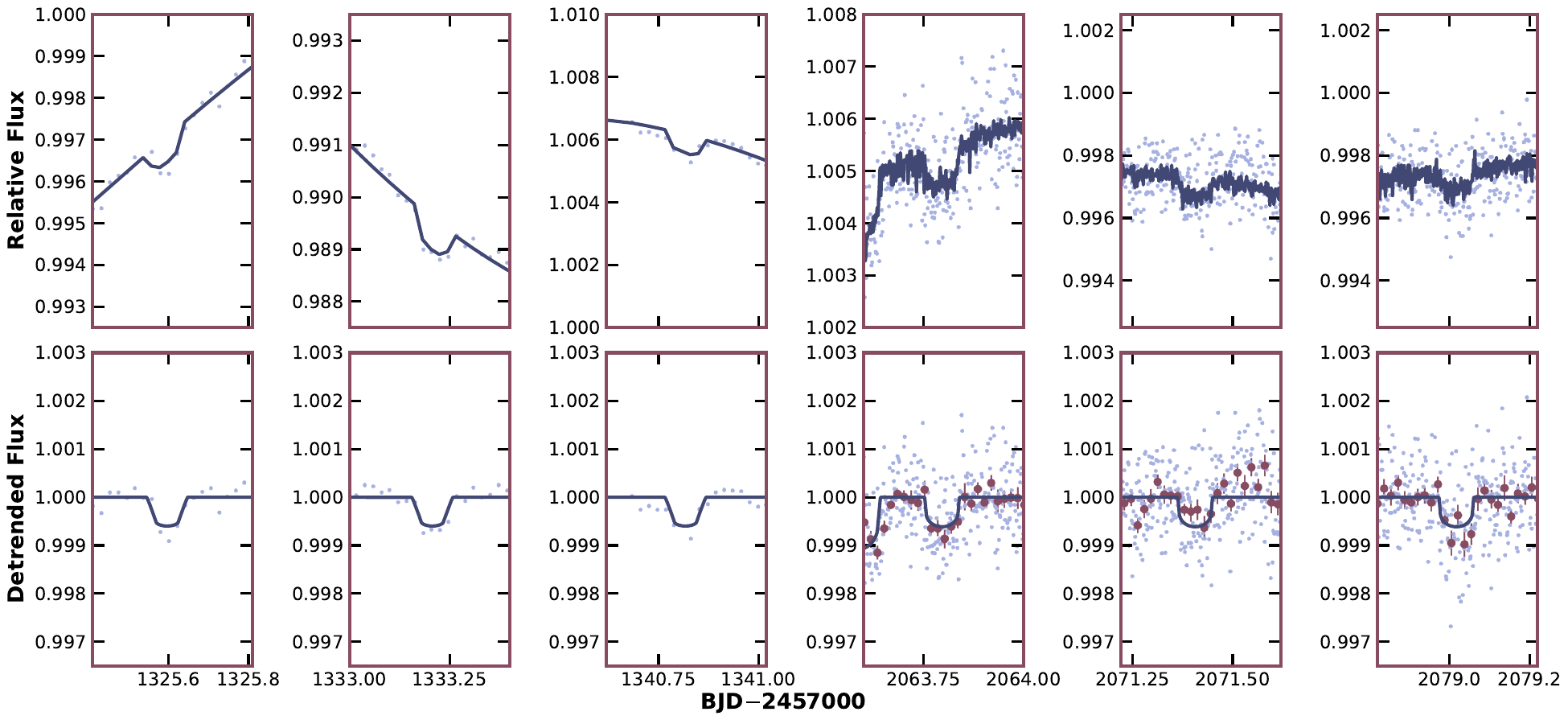}
    \caption{The \tess{} light curves centred on the transits of \targetb{}. The top panel shows the pre-detrending, the bottom panel shows the post-detrending light curves, after the removal of spot modulated rotational signal from \target{}. Columns 1-3 were observed at 30 minute cadence during Sector 1, while 4-6 were observed at 2 minute cadence from Sector 28. The best fit transit model is displayed in navy, and the detrended transits at 2 minute cadence have been binned in 10 minute intervals to illustrate the precision of \tess{}.}
    \label{fig:lc_zoom_01}
\end{figure*}

\begin{figure*}
    \centering
    \includegraphics[width=15cm]{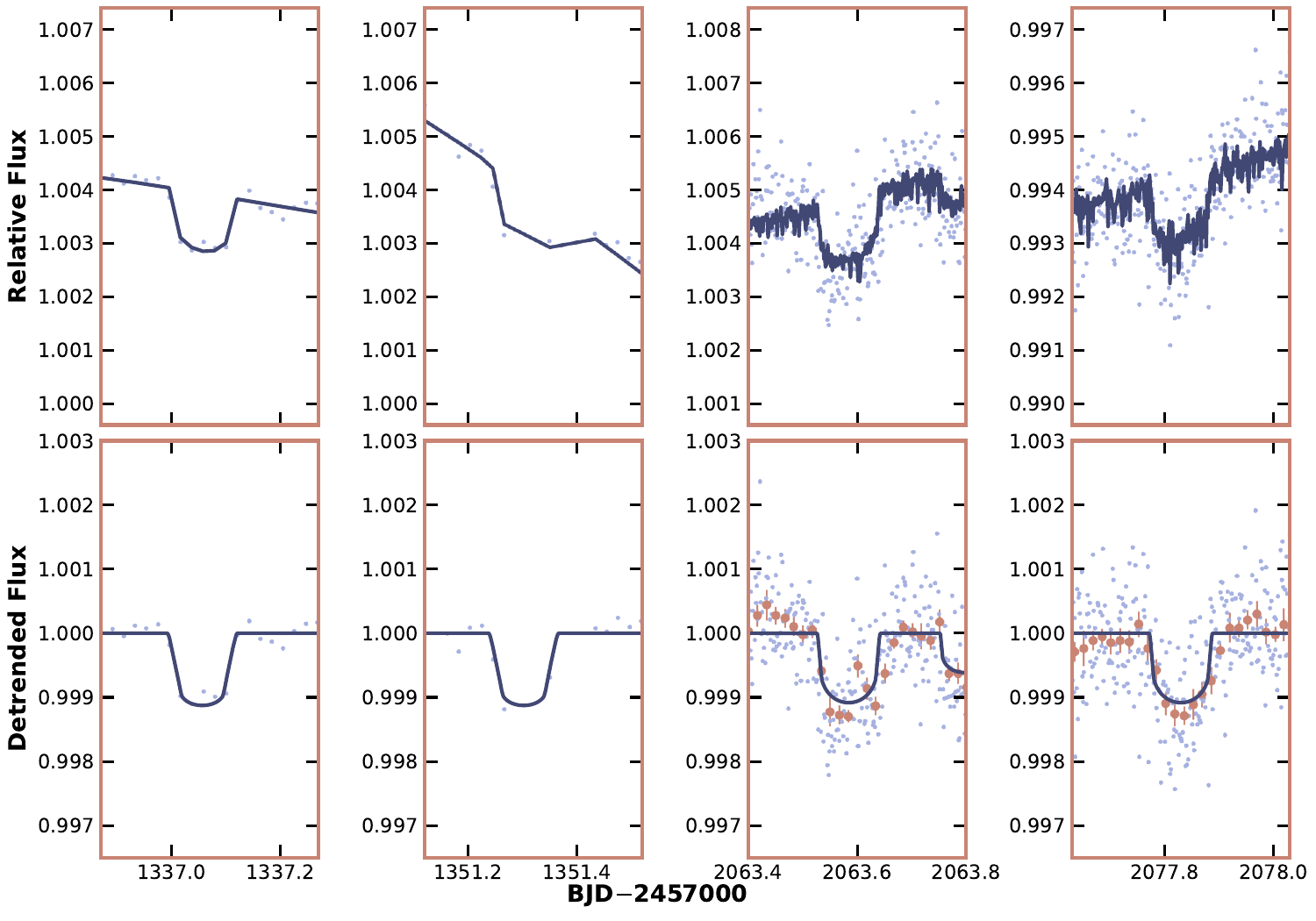}
    \caption{The \tess{} light curves centred on the transits of \targetc{}. The top panel shows the pre-detrending, the bottom shows the post-detrending light curve, after the removal of spot modulated rotational signal from \target{}. Columns 1-2 were observed at 30 minute cadence during Sector 1, while 3-4 were observed at 2 minute cadence during Sector 28. The best fit transit model is displayed in navy, and the detrended transits at 2 minute cadence have been binned in 10 minute intervals to illustrate the precision of \tess{}.}
    \label{fig:lc_zoom_02}
\end{figure*}

\begin{figure*}
    \centering
    \begin{tabular}{cc}
        \includegraphics[width=0.4\linewidth]{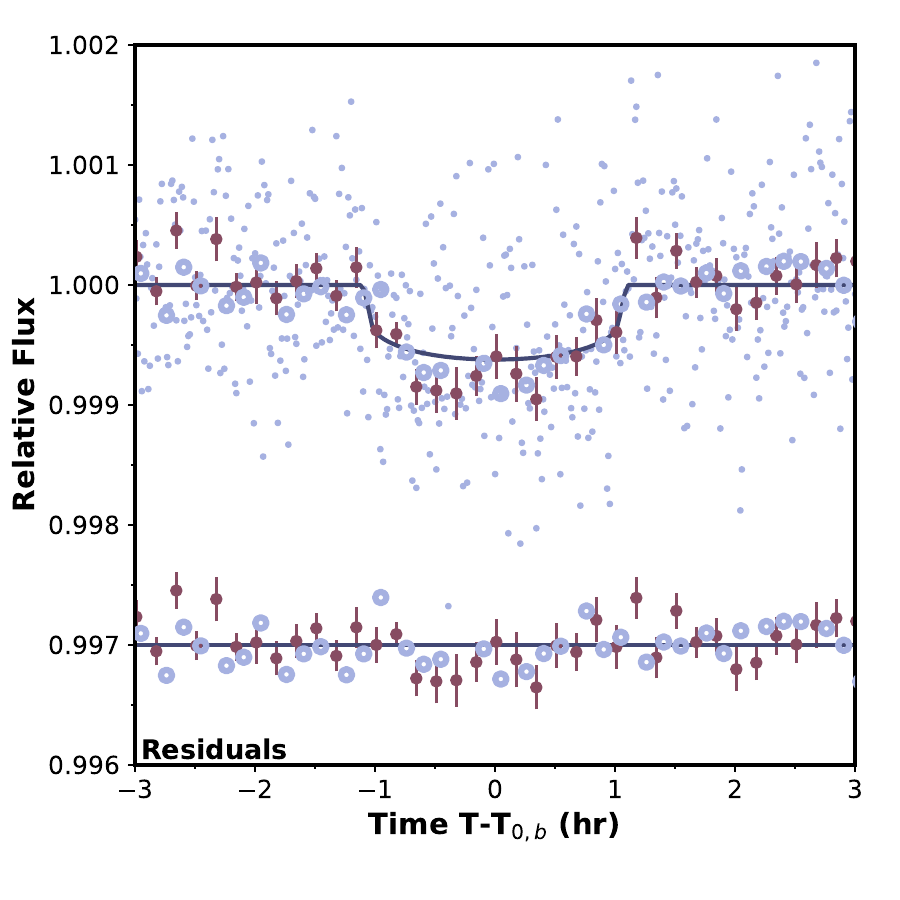} &
        \includegraphics[width=0.4\linewidth]{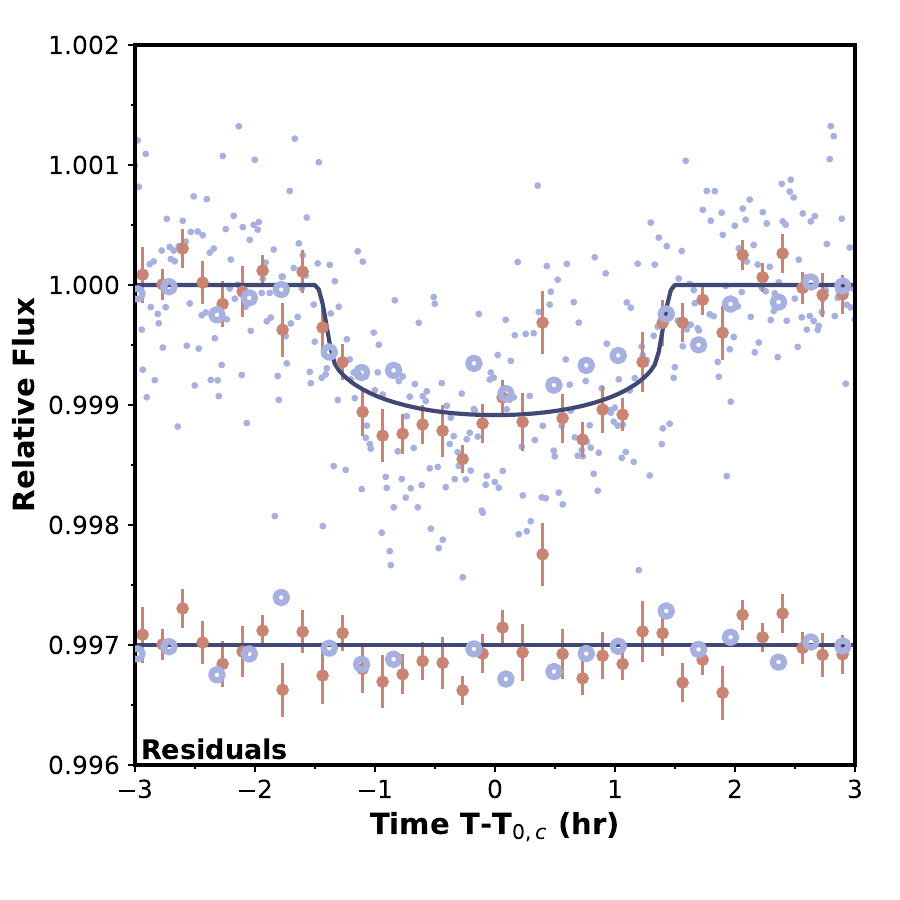} \\
    \end{tabular}
    \caption{Phase folded \textit{TESS} transit light curves for \targetb{} (Left) and \targetc{} (right). The light blue open circles represent 30 minute cadenced observations, while the light blue filled points represent the 2 minute cadenced observations. The points with error bars show the binned 2 minute light curves at 10 minute cadence. The best fit models are over plotted. The binned residuals are plotted at the bottom, vertically offset by 0.003 in flux for clarity. }
    \label{fig:phasefold}
\end{figure*}

\subsection{\cheops{} Follow-up Photometry} 
\label{sec:cheops}
Although we detected \targetb{} and \targetc{} through \tess{}, additional observations with higher precision are required to confirm and constrain the radius and ephemerides values for both planets. We therefore  use the \cheops{} mission to observe the primary transit of both planets during a single observing window. \cheops{} is a visible to infrared ($0.4\upmu\text{m} - 1.1\upmu\text{m}$) 0.32\,m Ritchey-Chretien telescope located in a 700\,km geocentric Sun-synchronous orbit. It is capable of capturing high-precision photometry of exoplanets around bright stars, with the corresponding \cheops mission focusing on the radius refinement of super-Earths and sub-Neptunes~\citep{2021_cheops}. 

The \cheops{} observation (observation ID: 1901592) was obtained between 2022 September 09 20:31 and 2022 September 10 14:06 UTC (10 orbits over $\sim$17.5 hours), with a $\sim$5 hour baseline between ingress and egress of both transits. At an exposure time of 60 seconds, 700 frames are obtained, with 10 frames affected by stray light and Earth occultation. This observation of a near-simultaneous transit for \targetb{} and \targetc{} was possible only because of the near 2:1 resonance of the system.

The low Earth orbit nadir-locked orientation of \cheops{} naturally induces field rotation over the course of a spacecraft-orbit, and results in correlated systematics in the observed light curve. We modelled these effects alongside the transit model as part of our global modelling (Section~\ref{sec:global_model}). The spacecraft signals are modelled as a linear combination of the sky background, smear, contamination, pixel X and Y drifts, and a set of four sine and cosine functions at frequencies up to four times the spacecraft orbital period as a function of the spacecraft roll angle. 

\begin{figure*}
    \centering
    \includegraphics[width=15cm]{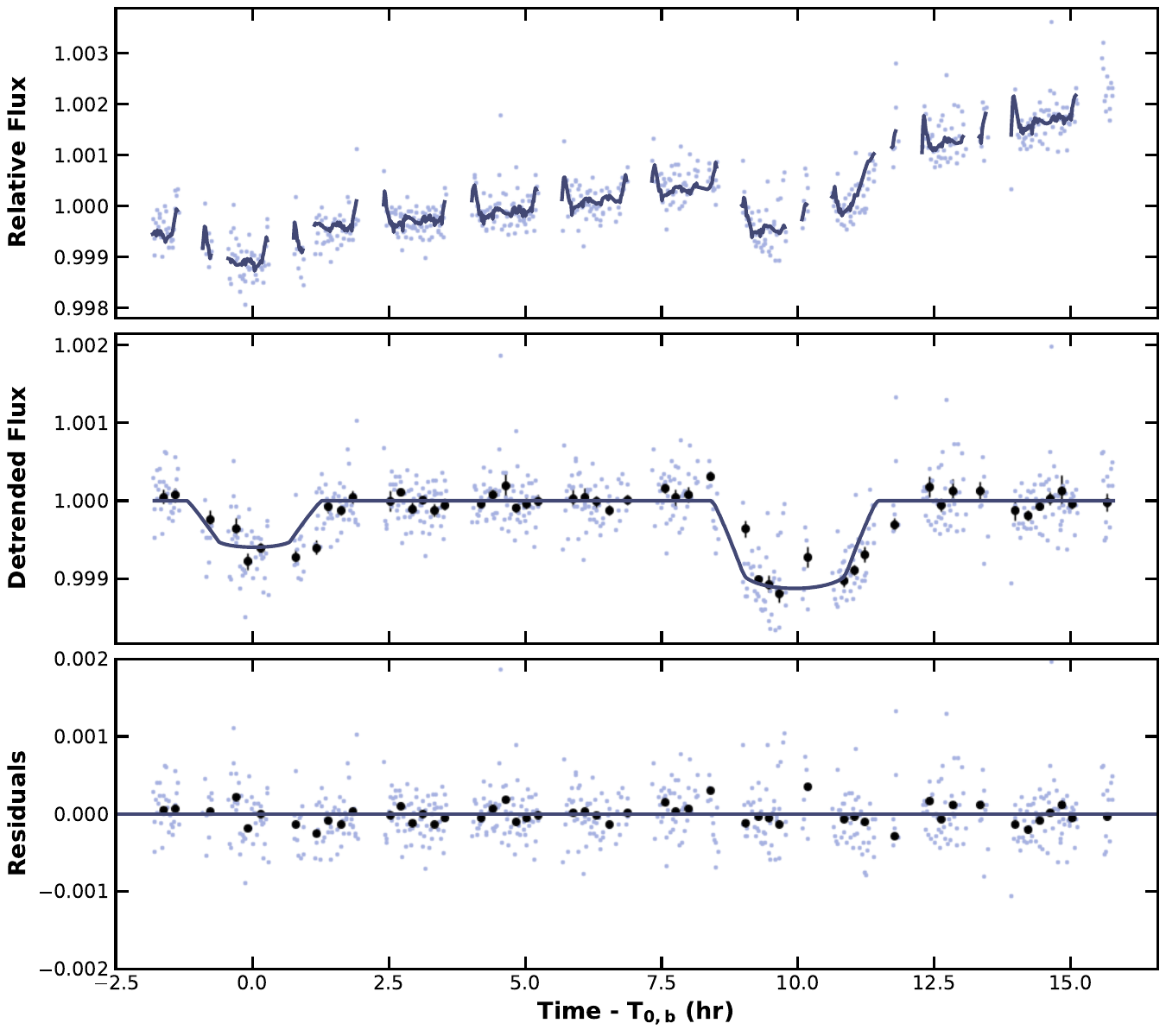}
    \caption{The follow-up \cheops{} observations of \targetb{} (first transit) and \targetc{} (second transit) taken over a single  $\sim$17.5 hour visit. The \textbf{top} panel displays the raw light curves from the optimal aperture extraction. The model describing the planetary transits and the instrumental effects is overplotted via the navy line (see Section~\ref{sec:global_model}). The \textbf{middle} panel shows the detrended CHEOPS light curve after removal of the instrumental spacecraft orbit induced variations and the best fit transit model. The transits are binned in 10 minute intervals to illustrate the precision of \cheops{}. The \textbf{bottom} panel illustrates the residuals of the data.}
    \label{fig:lc_cheops}
\end{figure*}

Figure \ref{fig:lc_cheops} shows the raw and detrended \cheops{} light curves, and the model describing the instrumental signals that were removed from the raw light curve.

\subsection{Ground Based Follow-up Photometry} 
\label{sec:ground_FU}

In addition to space-based observations, we also obtained ground-based seeing limited photometry through the \tess{} Follow-up Observing Program (TFOP) photometry science working group (SG1) to detect the transits of both planets and further rule out other nearby targets contaminating the detection.

\begin{figure*}
    \centering
    \includegraphics[width=15cm]{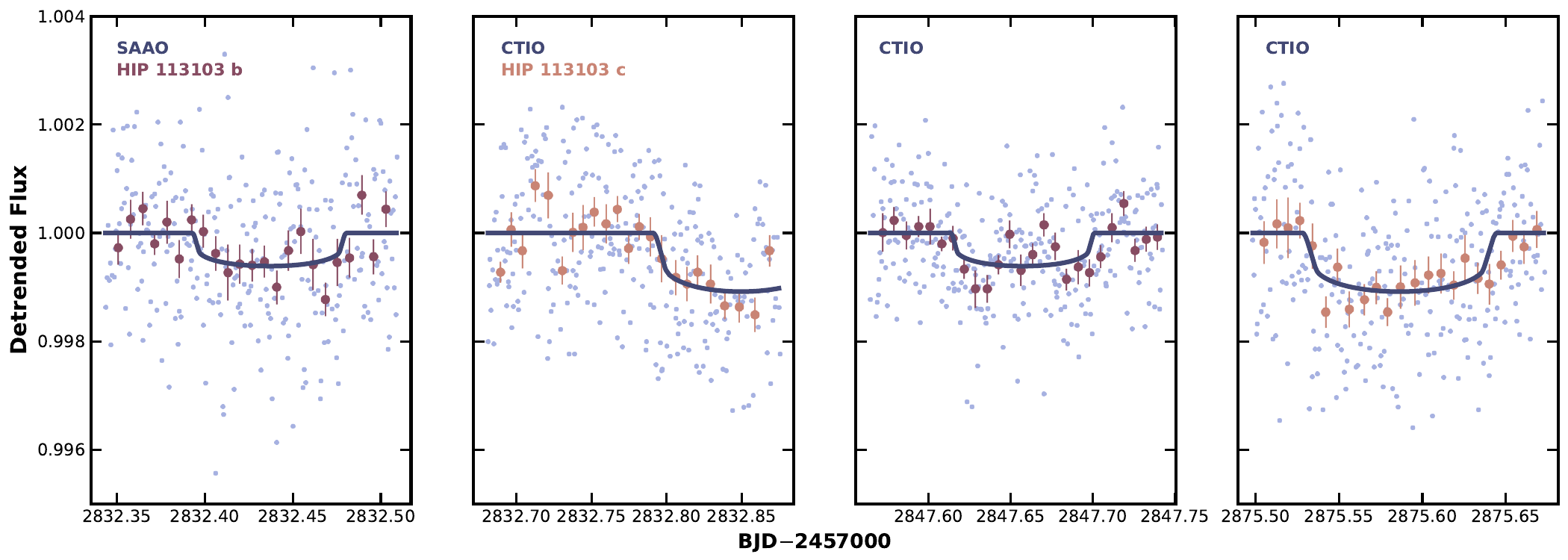}
    \caption{The ground-based detrended photometric follow-up observations of \targetb{} and \targetc{}, as obtained with the Las Cumbres Observatory 1\,m telescopes at SAAO and CTIO  (all in $z_s$ filter). The best fit transit model is represented via the navy line, while each transit has been binned in 10 minute intervals to illustrate the precision of the LCO telescopes.}
    \label{fig:lco}
\end{figure*}

Two transits of both \targetb{} and \targetc{} were obtained using the Las Cumbres Observatory Global Telescope \citep[LCOGT][]{2013PASP..125.1031B} facility. We used the 1\,m telescopes of the LCOGT network for these observations. Each telescope is equipped with a \emph{sinistro} $4\,\mathrm{K}\times 4\,\mathrm{K}$ andor EM CCD camera, yielding a field of view of $5.7'$ and a pixel scale of $0.34"\,\mathrm{pixel}^{-1}$. These observations are able to detect the transits of both planets with high significance, and determine that the transit depths are consistent with those derived from \tess{} and \cheops{}. The images were calibrated using the standard LCOGT {\tt BANZAI} pipeline \citep{McCully_2018} and the differential photometric data was extracted using {\tt AstroImageJ} \citep{Collins_2017}. Given \gaia{} DR3 catalog shows that no other stars are within $10"$ of \target{}, we determine that the transit signals most likely originated from the target star. The light curves are detrended simultaneously against the airmass in our global fit. Figure~\ref{fig:lco} shows the detrended light curves against their respective model light curve from our global model fit. The observations are detailed as follows:   

A full transit of \targetb{} was obtained via the 1\,m telescope at the South African Astronomical Observatory (SAAO) on UT 2022-09-09 with a 5.5$"$ radius aperture using the $z_s$ filter. On UT 2022-09-10, a partial transit of \targetc{}, including ingress, was obtained from the Cerro Tololo Interamerican Observatory (CTIO) node with a 6.2$"$ radius aperture using the $z_s$ filter. An additional full transit for both \targetb{} and \targetc were obtained from the CTIO node on UT 2022-09-10 and UT 2022-10-22. Both transits were obtained with the $z_s$ filter, with an aperture size of 8.2$"$ and 7.0$"$ for \targetb{} and \targetc{} respectively. Table~\ref{tab:obs_table} displays all the photometric transit observations analysed in this work.

\begin{table}
    \centering
    \caption{A summary of all ground-based photometric transit observations for \targetb{} and \targetc{}. These instruments are involved in the LCOGT consortium.}
    \label{tab:obs_table}
\begin{tabular}{ccccc}
    \hline \hline
    \textbf{Target} & \textbf{Instrument} & \textbf{Date (UT)} & \textbf{Filter} & \textbf{Aperture} \\ [0.3ex] 
    \hline
    \targetb{} & SAAO 1.0 m & 2022-09-09 & $z_{s}$ & 5.5'' \\
    \targetc{} & CTIO 1.0 m & 2022-09-10 & $z_{s}$ & 6.2'' \\
    \targetb{} & CTIO 1.0 m & 2022-09-25 & $z_{s}$ & 8.2'' \\
    \targetc{} & CTIO 1.0 m & 2022-10-22 & $z_{s}$ & 7.0'' \\
    \hline
\end{tabular}
\end{table}

\subsection{Spectroscopic Characterisation}
\label{sec:spec}
To characterise the stellar properties of the host star and validate the planetary-nature of the transiting candidates, we obtained a series of reconnaissance spectroscopic observations of \target{} with a set of southern spectroscopic facilities.

The stellar atmospheric parameters were derived by matching each observation against a library of $\sim$10,000 observed spectra previously classified through the Spectroscopic Classification Pipeline \citep{2012Natur.486..375B}. The library is interpolated via a gradient boosting regressor model, from which the best fit spectral parameters were determined \citep{2021AJ....161....2Z}. We found a best fit effective temperature of $T_\mathrm{eff} = 4930\pm100$\,K, surface gravity of $\log g = 4.6\pm0.1$ dex, metallicity of $\mathrm{[m/H]} = -0.1 \pm 0.1$ dex, and projected rotational broadening of $v\sin I_\star = 3\pm1\,\mathrm{km\,s}^{-1}$ for \target{}. We note that the rotational velocity is less than the instrument broadening, and the reported value is likely an upper limit of the true rotation velocity.

In addition, we obtained eight observations of \target{} using the CHIRON facility on the SMARTS 1.5-meter telescope located at Cerro Tololo Inter-American Observatory, Chile \citep{2013PASP..125.1336T}. CHIRON is a fibre-fed echelle spectrograph with a resolving power of $R \sim 80,000$ over the wavelength range of 4100 to $8700$\,\AA{}. We use the extracted spectra from CHIRON reduced via the standard pipeline as per \citet{2021AJ....162..176P}. The radial velocities are derived from each observation via a least-squares deconvolution of the spectra against a synthetic template generated at the atmospheric parameters of the target star \citep{1997MNRAS.291..658D}. The generated line profiles are modelled via a combination of kernels describing the rotational, macroturbulent, and instrument broadening effects \citep[following][]{1994AJ....107..742G}.

We also obtained ten epochs of spectroscopic observations from the \textsc{Minerva}-Australis array. \textsc{Minerva}-Australis is an array of four identical 0.7\,m telescopes, located at Mt Kent Observatory, Australia. The light from all four telescopes are combined into a single \textsc{KiwiSpec} high resolution echelle spectrograph, with a resolving power of $R\sim80,000$ over the wavelength region of $4800-6200$\,\AA{} \citep{2012_Barnes, 2019PASP..131k5003A}. Wavelength corrections are provided by two simultaneous fibers adjacent to the object fibers, which pass light from a quartz lamp through an iodine cell. Relative radial velocities are derived by a cross correlation between each individual observation and an averaged spectrum of the set of spectra available for the target. These relative velocities are then shifted to the mean absolute velocity of the averaged spectrum. These velocities are also presented in Table~\ref{tab:rv_table}.

The \textsc{Minerva}-Australis observations have per-point uncertainties of $10-20$\,\ms{}, and are comparable to those obtained from CHIRON. We do not detect significant radial velocity variations at the 20\,\ms{} level, consistent with the expected low mass of the planets around \target{}. The observations therefore remain consistent with a lack of detection of the radial velocity orbit, as is expected given the velocity uncertainties and the expected orbit amplitude. The line profiles exhibit no visible variations indicative of blend scenarios. In scenarios where the transit is induced by a background eclipsing binary, we would often observe correlations between the rotational broadening velocity and the radial velocities, with the apparent broadening at its maximum at the extremities of the velocity curve. We observe no such correlation for \target{}, with the exposure to exposure scatter in the rotational broadening of 0.2\,\kms{}. 

In addition, two archival spectra were obtained from the European Southern Observatory (ESO) HARPS facility on the ESO 3.6\,m telescope in La Silla, Chile~\citep{2003_Mayor}. The observations have a spectral resolution of $R=120,000$ over the spectral range of $3780-6910$\,\AA{}. We make use of the two archival observations, obtained in 2010 and 2013, to further classify the host star atmospheric properties. To calculate the spectroscopic parameters, we make use of the \texttt{ZASPE} package \citep{2017MNRAS.467..971B} and its associated custom spectral library computed from the \citet{Castelli:2004} model atmospheres. We find a mean effective temperature of  $T_\mathrm{eff} = 4800\pm60$\,K, surface gravity of $\log g = 4.47\pm0.05$ dex, and metallicity of $\mathrm{[m/H]} = 0.0 \pm 0.05$ dex, with uncertainties adapted from the uncertainty floor as per \citet{2017MNRAS.467..971B}. We do not incorporate the HARPS observations towards our spectroscopic parameters due to the sample being too small. We instead adopt the CHIRON and \textsc{Minerva}-Australis spectra for our spectroscopic parameters, as we were able to test the self consistency of our parameters via its scatter from spectrum to spectrum (as presented in Table~\ref{tab:stellar_parameters}).

In addition, as the HARPS spectra cover the Calcium H \& K lines, we also make use of the two available spectra to compute activity indices for \target{}. We followed the same procedure as per the Mt Wilson catalog \citep{1978ApJ...226..379W,1978PASP...90..267V,1991ApJS...76..383D,1995ApJ...438..269B}, and compute the S-index via a set of photometric band passes about the line cores and continuum around each line. The S-index is then converted to the $\log R'_\mathrm{HK}$ index as per \citet{1984ApJ...279..763N}. We found a mean Calcium H\,K activity of $\log R'_\mathrm{HK} = -4.69\pm0.05$ from the two HARPS observations, indicating minimal chromospheric activity being exhibited by the host star~\citep{1996_Henry}. 

\begin{table}
    \centering
    \caption{Radial velocity measurements of \target{}}
    \label{tab:rv_table}
\begin{tabular}{cccc}
    \hline \hline
    \textbf{BJD} & \textbf{RV (km s$^{\mathbf{-1}}$)} & \textbf{$\mathbf{\sigma}_\text{RV}$ (km s$^{\mathbf{-1}}$)} & \textbf{Instrument}\\ [0.3ex] 
    \hline
    2459171.60440 & 12.813 & 0.035 & CHIRON\\
    2459174.59126 & 12.864 & 0.024 & CHIRON\\
    2459176.58998 & 12.824 & 0.022 & CHIRON\\
    2459178.62741 & 12.858 & 0.026 & CHIRON\\
    2459180.55586 & 12.818 & 0.029 & CHIRON\\
    2459182.53304 & 12.830 & 0.034 & CHIRON\\
    2459184.55875 & 12.854 & 0.026 & CHIRON\\
    2459186.62126 & 12.877 & 0.022 & CHIRON\\
    2459917.93684 & 13.370 & 0.022 & \textsc{Minerva}-Australis\\
    2459917.95510 & 13.425 & 0.019 & \textsc{Minerva}-Australis\\
    2459924.93091 & 13.377 & 0.015 & \textsc{Minerva}-Australis\\
    2459924.94916 & 13.382 & 0.020 & \textsc{Minerva}-Australis\\
    2459930.93397 & 13.356 & 0.038 & \textsc{Minerva}-Australis\\
    2459930.95226 & 13.406 & 0.020 & \textsc{Minerva}-Australis\\
    2459942.92801 & 13.384 & 0.019 & \textsc{Minerva}-Australis\\
    2459942.94630 & 13.407 & 0.018 & \textsc{Minerva}-Australis\\
    2460046.29239 & 13.435 & 0.035 & \textsc{Minerva}-Australis\\
    2460046.31065 & 13.409 & 0.014 & \textsc{Minerva}-Australis\\
    \hline
\end{tabular}
\end{table}

\section{Global model} 
\label{sec:global_model}
In order to constrain the stellar and planetary properties of the HIP 113103 system, we performed a global model fit using all the observations outlined in Section~\ref{sec:obs}. Our global model is similar to that implemented in previous papers~\citep[e.g.][]{2022AJ....163..289Z}, and was tested against other publicly available codes such as \texttt{EXOFASTv2} in~\citet{2017_Rodriguez}. Free parameters fitted for include orbital parameters defining the orbital eccentricity \ecos{} and \esin{} (where $e$ is eccentricity and $\omega$ is argument of periapsis), line of sight inclination $i$, orbital period $P$, radius ratio \Rp{}/\rstar{}, and transit midpoint $T_0$. The photometric transits are modeled via \texttt{batman}~\citep{2015_Kreidberg} as per \citet{2002_Mandel}, simultaneously incorporating an associated instrument model to account for additional variability induced by external factors. This includes fitting for a polynomial accounting for the influence of spacecraft on the photometric fluxes for \cheops{} as per \citet{2022MNRAS.514...77M}, described in Section~\ref{sec:cheops}. Similarly, we also fit for linear coefficients to the mean, standard deviation, and skew terms of the three quarternions for \tess{} as per \citet{2019ApJ...881L..19V}. Ground based LCO photometry were simultaneously detrended against airmass to remove hours timescale variability in the baseline. Limb darkening coefficients are interpolated from the CHIRON stellar atmospheric parameters for the host star via \citet{Claret:2011} and \citet{2017AA...600A..30C}, and constrained by Gaussian priors with widths of 0.02. The width of the Gaussian prior is set by the uncertainties in the models, and by the propagated uncertainties from the spectroscopically derived stellar parameters. We also trialled the same global model, but with Gaussian priors of width 0.1 for the limb darkening parameters and note no significant changes to our model posteriors. Supersampling corrections of the light curve model has been applied where necessary when modelling the 30 minute cadenced observations~\citep{2010_Kipping}. The CHIRON and \textsc{Minerva}-Australis radial velocities were modeled via the \texttt{RadVel} package \citep{2018PASP..130d4504F}, accounting for their respective instrumental offsets and velocity jitter terms. 

To jointly model the stellar properties, we interpolate the MIST isochrones~\citep{2016_Dotter} along age, stellar mass, and metallicity, with outputs of stellar radius and absolute magnitudes in a set of photometric bands as is made available by the public isochrone files. The spectral energy distribution and \emph{Gaia} parallax provide the tightest observational constraints on the host star properties. At each iteration, we include jump parameters for age, host star mass $M_\star$, metallicity [M/H], and parallax. The parallax of the target is strongly constrained by a Gaussian prior about that measured by \gaia{} DR3 \citep{2022arXiv220800211G}, with a correction of $-0.025657$\,mas applied according to \citet{2021A&A...649A...4L}. We compare the interpolated MIST absolute magnitudes against that of the observed  \hipp{} TYCHO $B$, and $V$, 2MASS $J$, $H$, and $K$, and the \gaia{} $G$, $B_{p}$, and $R_{p}$ bands~\citep{1997_Perryman, 2006_Skrutskie, 2018_Gaia} magnitudes, after correcting for the distance modulus via the parallax jump parameter. In addition to the absolute magnitudes, we also interpolate the MIST isochrones along stellar radius, which is then incorporated into modelling of the transit parameters, such as $a/R_\star$. 

We constrained our models using a Markov Chain Monte Carlo analysis via \texttt{emcee}~\citep{emcee}, with 400 walkers over 4000 iterations per walker (with the first 2000 iterations allocated to burn in). Informative priors are summarised in Table~\ref{tab:planet_parameters}, while all other fitted parameters are constrained by uniform priors bounded by their physical limits. The derived planetary and stellar values are presented in Table~\ref{tab:planet_parameters} and Table~\ref{tab:stellar_parameters} respectively. Figure~\ref{fig:lc_tess} shows our output model for our \tess{} dataset, Figure~\ref{fig:lc_cheops} for \cheops{}, and Figure~\ref{fig:lco} for ground based photometric follow-up observations.

\begin{table*}
    \centering
    \caption{The physical properties of \target{}}
    \label{tab:stellar_parameters}
\begin{tabular}{|l|r|p{4cm}|}
    \hline \hline
    \textbf{Parameter} & \textbf{Value} & \textbf{Source}\\ 
    \hline
    ~~~\textbf{Identifiers} \dotfill  & HIP 113103 & \\
    & TIC 121490076 & \\
    & TYC 8011-00766-1 & \\
    & 2MASS J22541736-4300372 & \\
    & Gaia DR2 6541360574788758016 & \\
    \multicolumn{3}{l}{\textbf{Astrometry}\vspace{2mm}} \\
    ~~~Right Ascension \dotfill  & \ra{} & \citet{2022arXiv220800211G}\\
    ~~~Declination \dotfill & \dec{} & \citet{2022arXiv220800211G}\\
    ~~~Parallax (mas) \dotfill & \parallax{} & \citet{2022arXiv220800211G}\vspace{4mm}\\
    \multicolumn{3}{l}{\textbf{Proper Motion}\vspace{2mm}} \\
    ~~~Gaia (2016.4) RA Proper Motion (\masyr{}) \dotfill & \gaiapmra{} & \citet{2022arXiv220800211G}\\
    ~~~Gaia (2016.3) DEC Proper Motion (\masyr{}) \dotfill & \gaiapmdec{} & \citet{2022arXiv220800211G}\\
    ~~~Hipparcos (1991.2) RA Proper Motion (\masyr{}) \dotfill & \hippara{} & \citet{1997_Perryman}\\
    ~~~Hipparcos (1991.4) DEC Proper Motion (\masyr{}) \dotfill & \hippadec{} & \citet{1997_Perryman}\\
    ~~~Hipparcos-Gaia Average RA Proper Motion (\masyr{}) \dotfill & \gavgpmra & \citet{2021_Brandt}\\
    ~~~Hipparcos-Gaia Average DEC Proper Motion (\masyr{}) \dotfill & \gavgpmdec & \citet{2021_Brandt}\vspace{2mm}\\
    \multicolumn{3}{l}{\vspace{2mm}\textbf{Photometry}} \\
    ~~~\tess{} (mag) \dotfill & \tessmag & \citet{2019_Stassun}\\
    ~~~B (mag) \dotfill & \bmag{} & \citet{2000AA...355L..27H}\\
    ~~~V (mag) \dotfill & \vmag{} & \citet{2000AA...355L..27H}\\
    ~~~J (mag) \dotfill & \jmag{} & \citet{2006_Skrutskie}\\
    ~~~H (mag) \dotfill & \hmag{} & \citet{2006_Skrutskie}\\
    ~~~K (mag) \dotfill & \kmag{} & \citet{2006_Skrutskie}\\
    ~~~\gaia{} $G$ (mag) \dotfill & \gmag{} & \citet{2022arXiv220800211G}\\
    ~~~\textit{Gaia}$_{BP}$ (mag) \dotfill & \gmagbp{} & \citet{2022arXiv220800211G}\\
    ~~~\textit{Gaia}$_{RP}$ (mag) \dotfill & \gmagrp{} & \citet{2022arXiv220800211G}\\
    ~~~\wise{} W1 (mag) \dotfill & \wonemag{} & \citet{2012_Cutri}\\
    ~~~\wise{} W2 (mag) \dotfill & \wtwomag{} & \citet{2012_Cutri}\\
    ~~~\wise{} W3 (mag) \dotfill & \wthreemag{} & \citet{2012_Cutri}\\
    ~~~\wise{} W4 (mag) \dotfill & \wfourmag{} & \citet{2012_Cutri}\vspace{2mm}\\
    \multicolumn{3}{l}{\textbf{Kinematics and Position}\vspace{2mm}} \\
    ~~~$U$ (\kms{}) \dotfill & \ukms{} & Propagated from Gaia$^1$\\
    ~~~$V$ (\kms{}) \dotfill & \vkms{} & Propagated from Gaia$^1$\\
    ~~~$W$ (\kms{}) \dotfill & \wkms{} & Propagated from Gaia$^1$\\
    ~~~Distance (pc) \dotfill & \dstar{} & This paper\vspace{1mm}\\
    ~~~$\gamma_{\mathrm{CHIRON}}$ (\kms{}) \dotfill & $12.845_{-0.013}^{+0.012}$ & This paper\vspace{1mm}\\
    ~~~$\gamma_{\mathrm{MINERVA}}$ (\kms{}) \dotfill & $13.395_{-0.011}^{+0.010}$ & This paper\vspace{1mm}\\
    ~~~$\mathrm{Jitter}_{\mathrm{CHIRON}}$ (\ms{}) \dotfill & $16_{-11}^{+20}$ & This paper\vspace{1mm}\\
    ~~~$\mathrm{Jitter}_{\mathrm{MINERVA}}$ (\ms{}) \dotfill & $13_{-9}^{+14}$ & This paper\vspace{2mm}\\
    \multicolumn{3}{l}{\textbf{Physical Properties}\vspace{2mm}} \\
    ~~~\mstar{} (\msun) \dotfill & \mhstar{} & This paper\\
    ~~~\rstar{} (\rsun) \dotfill & \rhstar{} & This paper\\ 
    ~~~\teff{} (K) \dotfill & \teffstar{} & This paper\\
    ~~~\logg{} (cgs) \dotfill & \loggstar{} & This paper\\
    ~~~\m{} \dotfill & \metalicity{} & This paper\\
    ~~~\vsini{} (\kms{}) \dotfill & \vsinistar{} & This paper\\
    ~~~Rotation period (d) \dotfill & $9.92\pm0.23$ & This paper\\    
    ~~~Gyrochronology Age (Myr) \dotfill & \age{} & Based on the gyrochronology relationship from \citet{Bouma_2023}\\
    ~~~Limb darkening coefficients (\tess{}$_\mathrm{u1}$) \dotfill & $0.463\pm0.021$ & \citet{2017AA...600A..30C}\\
    ~~~Limb darkening coefficients (\tess{}$_\mathrm{u2}$) \dotfill & $0.182\pm0.020$ & \citet{2017AA...600A..30C}\\
    ~~~Limb darkening coefficients (\cheops{}$_\mathrm{u1}$) \dotfill & $0.604\pm0.021$ & \citet{Claret:2011} \\
    ~~~Limb darkening coefficients (\cheops{}$_\mathrm{u2}$) \dotfill & $0.111\pm0.022$ & \citet{Claret:2011}\\
    ~~~Limb darkening coefficients (LCO z' band $_\mathrm{u1}$) \dotfill & $0.350\pm0.021$ & \citet{Claret:2011}\\
    ~~~Limb darkening coefficients (LCO z' band $_\mathrm{u2}$) \dotfill & $0.287\pm0.021$ & \citet{Claret:2011}\\
    \hline
\end{tabular}\\
\footnotesize{$^1$ Propagated from \textit{Gaia} via the \textsc{gal uvw} function in the \textsc{PyAstronomy} package \citep{pya}.}
\end{table*}

\begin{table*}
    \centering
    \renewcommand{\arraystretch}{1.3}
    \caption{Derived parameters for HIP 113103 b and HIP 113103 c. Values denoted with an asterisk were calculated using an estimated mass derived from the method outlined in~\citet{2016_Wolfgang}, as described in Section~\ref{sec:planet_prop}. For \Teq{}, we assume a $A_{B}=0$.}
    \label{tab:planet_parameters}
\begin{tabular}{lrr}
    \hline \hline
    \textbf{Parameter} & \textbf{Value} & \textbf{Prior}\\ 
    \hline
    \multicolumn{3}{l}{\textbf{Fitted parameters for \targetb{}}\vspace{2mm}} \\
    ~~~$T_0$ (BJD-TDB) \dotfill & \tcb{} & Fitted\\
    ~~~$P$ (days) \dotfill & \periodb{} & Fitted\\
    ~~~\RpRs{} (\rstar{}) \dotfill & \rprsb{} & Fitted\\
    ~~~$i$ (deg) \dotfill & \incb{}  & Fitted\\
    ~~~\ecos{} \dotfill & \ecosomegab{} & Fitted\\
    ~~~\esin{} \dotfill & \esinomegab{} & Fitted\vspace{4mm}\\
    \multicolumn{3}{l}{\textbf{Inferred parameters for \targetb{}}} \\
    ~~~$e$ \dotfill & \eccentricityb{} & Derived\\
    ~~~$\omega$ (deg) \dotfill & \omegab{} & Derived\\
    ~~~\Rp{} (\rearth{}) \dotfill & \rpb{} & Derived\\
    ~~~\aRs{} (\rstar{}) \dotfill & \abvsrb{} & Derived\\
    ~~~$a$ (AU) \dotfill & \ab{} & Derived\\
    ~~~\Tdur{} (days) \dotfill & \tfourteenb{} & Derived\\
    ~~~\Teq{} (K) \dotfill & \teqb{} & Derived\\
    ~~~\imppar{} \dotfill & \bb{} & Derived\vspace{1mm}\\
    ~~~(\RpRs{})$^{2}$ \dotfill & \rprsbsq{} & Derived \\
    ~~~\Mp{} (\Me{}) from mass-radius relationships \dotfill & 5.9 $\pm$1.9$^{*}$ & Inferred \\
    ~~~\Krv{} (\ms{}) from mass-radius relationships \dotfill & \kvb{} & Inferred\\
    ~~~\rhop (\rhoe{}) from mass-radius relationships \dotfill & \rhob{}$^{*}$ & Inferred\vspace{4mm}\\
    \multicolumn{3}{l}{\textbf{Fitted parameters for \targetc{}}} \\
    ~~~$T_0$ (BJD-TDB) \dotfill & \tcc{} & Fitted\\
    ~~~$P$ (days) \dotfill & \periodc{} & Fitted\\
    ~~~\RpRs{} (\rstar{}) \dotfill & \rprsc{} & Fitted\\
    ~~~$i$ (deg) \dotfill & \incc{} & Fitted\\
    ~~~\ecos{} \dotfill & \ecosomegac{} & Fitted\\
    ~~~\esin{} \dotfill & \esinomegac{} & Fitted\vspace{4mm}\\
    \multicolumn{3}{l}{\textbf{Inferred parameters for \targetc{}}} \\
    ~~~$e$ \dotfill & \eccentricityc{} & Derived\\
    ~~~$\omega$ (deg) \dotfill & \omegac{} & Derived\\
    ~~~\Rp{} (\rearth{}) \dotfill & \rpc{} & Derived\\
    ~~~\aRs{} (\rstar{}) \dotfill & \acvsrc{} & Derived\\
    ~~~$a$ (AU) \dotfill & \ac{} & Derived\\
    ~~~\Tdur{} (days) \dotfill & \tfourteenc{} & Derived\\
    ~~~\Teq{} (K) \dotfill & \teqc{} & Derived\\
    ~~~\imppar{} \dotfill & \bc{} & Derived\vspace{1mm}\\
    ~~~(\RpRs{})$^{2}$ \dotfill & \rprscsq{} & Derived\\
    ~~~\Mp{} (\Me{}) from mass-radius relationships \dotfill & $8.4 \pm1.9^{*}$ & Inferred\\
    ~~~\Krv{} (\ms{}) from mass-radius relationships \dotfill & \kvc{} & Inferred\\
    ~~~\rhop (\rhoe{}) from mass-radius relationships \dotfill & \rhoc{}$^{*}$ & Inferred\vspace{4mm}\\
    \hline
\end{tabular}
\end{table*}

\begin{figure}
    \centering
    \includegraphics[width=0.45\textwidth]{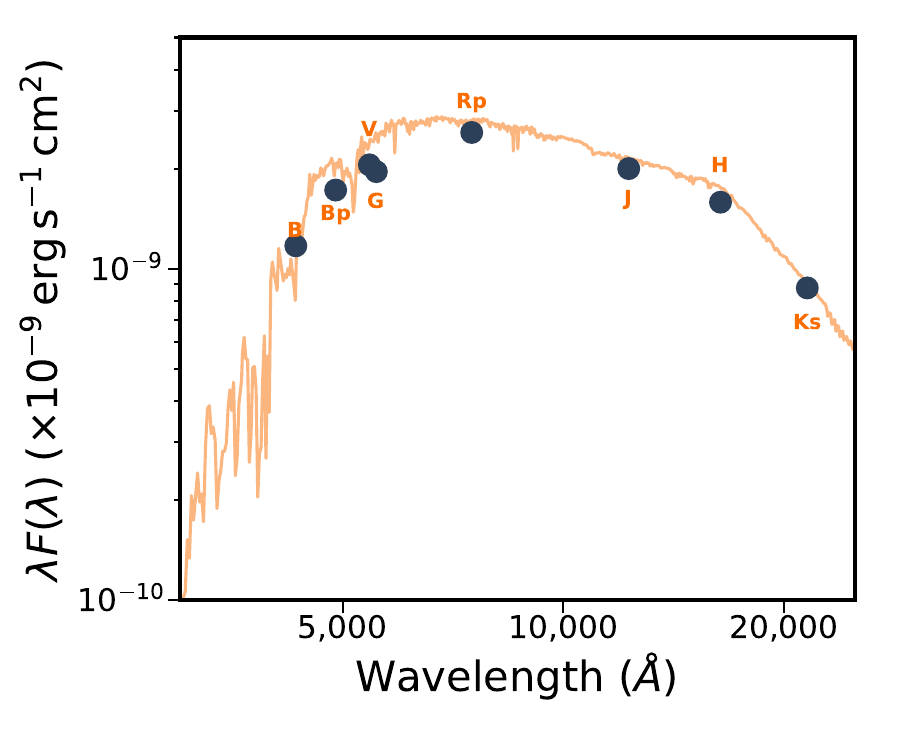}
    \caption{Spectral energy distribution of \target{}. We make use of the spectroscopic atmospheric priors and the photometric magnitudes of \target{} to constrain the stellar properties simultaneous to our global modeling of the stellar and planetary parameters. }
    \label{fig:SED}
\end{figure}

\section{Stellar Rotation and Age}
\label{sec:stellar}

\begin{figure}
    \centering
    \includegraphics[width=\columnwidth]{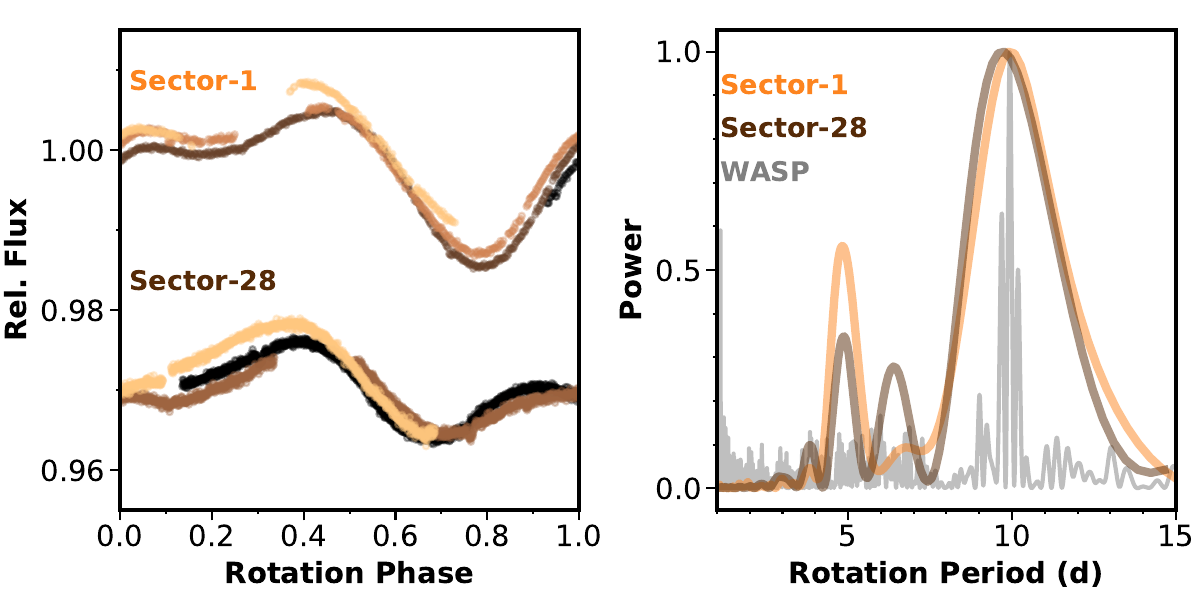}
    \caption{\target{} exhibits significant spot-induced rotational variability in its light curves. \textbf{Left} \tess{} light curves from Sectors 1 and 28 folded to the rotation period of \target{}; each rotation cycle is plotted in a progressively lighter shade. The sectors are separated by an arbitrary vertical offset. \target{} maintains a constant rotation period over the multi-year observations obtained by \tess{}. \textbf{Right} Autocorrelation periodograms of the \tess{} and SuperWASP light curves of \target{}, showing a consistent peak at 10.0 days over the course of more than 10 years.}
    \label{fig:lc_rotation}
\end{figure}

The \tess{} light curve of \target{} exhibits significant quasi-periodic variability at the 0.5\% level representative of rotational variability. Figure~\ref{fig:lc_rotation} shows the auto-correlation function of the periodicity over the two \tess{} sectors. We found a rotation period of $10.2 \pm 1.4$ days from Sector 1, and $10.0 \pm 1.3$ days from Sector 28 observations. The uncertainties were estimated by taking the width of the best fit Gaussian to the periodogram peaks. The rotation period is consistent between the two sectors, spanning one year in separation. 

In addition, \target{} also received one year of observations from the Wide Angle Search for Planets (WASP) Consortium \citep{2006PASP..118.1407P} with the Southern SuperWASP facility, located at the Sutherland Station of the South African Astronomical Observatory. The SuperWASP observatory consists of eight Canon 200\,mm f/1.8 telephoto lenses, yielding a $7.8^\circ\,\times\,7.8^\circ$ field of view each over a $2\mathrm{K}\,\times\,2\mathrm{K}$ detector. SuperWASP observations of \target{} spanned 2006-05-07 to 2007-11-13, yielding $\sim$11,000 epochs of observations. The periodogram from the SuperWASP light curves are also overplotted in Figure~\ref{fig:lc_rotation}, yielding a rotation period of $9.90\pm0.23$ days, consistent with the \tess{} observations more than a decade later. When combined, the TESS and WASP datasets provide a long term stable rotation period of $9.92 \pm 0.23$ days for \target{}. In addition, we make use of the measured rotational velocity $v\sin I_\star$ and the photometric rotation period to provide a $1\sigma$ lower limit for the stellar inclination angle of $I_\star > 56^\circ$ \citep{2020AJ....159...81M}, consistent with an aligned geometry. Using $R_{\star}=0.742\pm0.013\,R_{\odot}$ and $P_{\text{rot}}=9.92\pm0.23$ days, we also calculate an equatorial velocity of $V_{\text{eq}}=3.78\pm0.11$ km s$^{-1}$, which is in good agreement with our $v\,\text{sin}\,i$ value of $3\pm1$ km s$^{-1}$.

The rotation period of \target{} is consistent with an adolescent K dwarf. We adopt the rotation sequence interpolation presented by \citet{Bouma_2023}, and derive a rotation-based age of $470_{-110}^{+170}$ Myr at $1\sigma$ significance. Similarly, based on the rotation age relationship from \citet{2008ApJ...687.1264M}, the $1\sigma$ age range for \target{} is 380-510 Myr. However, gyrochronology is particularly insecure in estimating the ages of single K dwarfs. The spins of these stars may stall within the first billion years, and many around giga-year clusters exhibit similar rotation periods \citep[e.g.][]{2009ApJ...695..679M,2018ApJ...862...33A,2019ApJ...879..100D}. \citet{2015MNRAS.450.1787A} accounts for a larger spread in the spin-down dispersion of low mass stars, and the relationship they provide yields a $1\sigma$ age range of 200-2000 Myr for \target{}. \target{} lacks spectroscopic features, such as Li 6708\,\AA{} absorption and significant Calcium II H\&K emission that are usually indicative of youth, as is expected for K dwarfs older than $\sim$300 million years. The Calcium II H\&K derived index $\log R'_\mathrm{HK} = -4.69\pm0.05$ corresponds to an age of $1.9_{-0.5}^{+0.7}$ Gyr, consistent with the rotational derived age estimate. In addition, the isochrone modelling also provides a loose age constraint of $5\pm2$ Gyr at the $1\sigma$ level. We find no evidence that \target{} is kinematically associated with comoving stars via the \texttt{comove} package \citep{2021AJ....161..171T}\footnote{\url{https://github.com/adamkraus/Comove}}. It is therefore difficult to confirm the suspected youth of \target{}.

\begin{figure*}
    \centering
    \includegraphics[width=12cm]{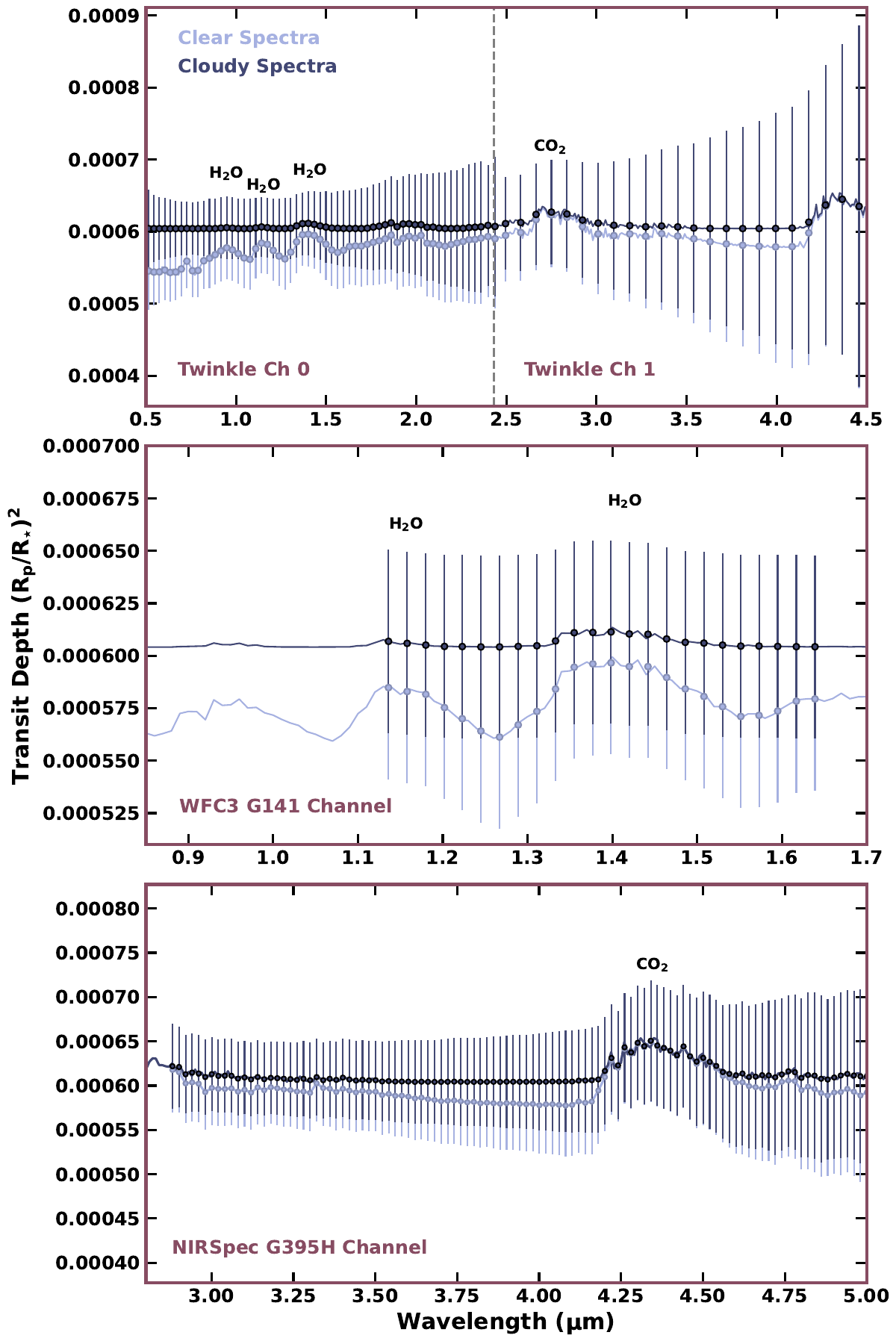}
    \caption{A synthetic atmospheric transmission spectra of \targetb, incorporating absorbing species CH$_{4}$, CO, CO$_{2}$, H$_{2}$O, and NH$_{3}$ at chemical equilibrium assuming an atmospheric metallicity of $100\,\times$ Solar. The \textbf{top} panel illustrates the expected spectrum after 10 visits with \twinkle{}, the \textbf{middle} panel shows the simulated spectrum from a single visit with \hst{} WFC3 with the G141 grism, and the \textbf{bottom} panel shows the simulated spectrum from a single visit with \jwst{} NIRSpec G395H grism (with each instrument capturing a different wavelength range). For each facility, we have illustrated two scenarios: one where spectrum is cloud-free ($10^{6}$ Pa), and another with a grey cloud deck ($10^{1}$ Pa). The synthetic spectra for \twinkle{} was obtained using the radiometric tool on the \twinkle{} Stardrive portal, while the \hst{} and \jwst{} spectra were generated using \texttt{PandExo}.}
    \label{fig:tuarex_hip113103b}
\end{figure*}

\begin{figure*}
    \centering
    \includegraphics[width=12cm]{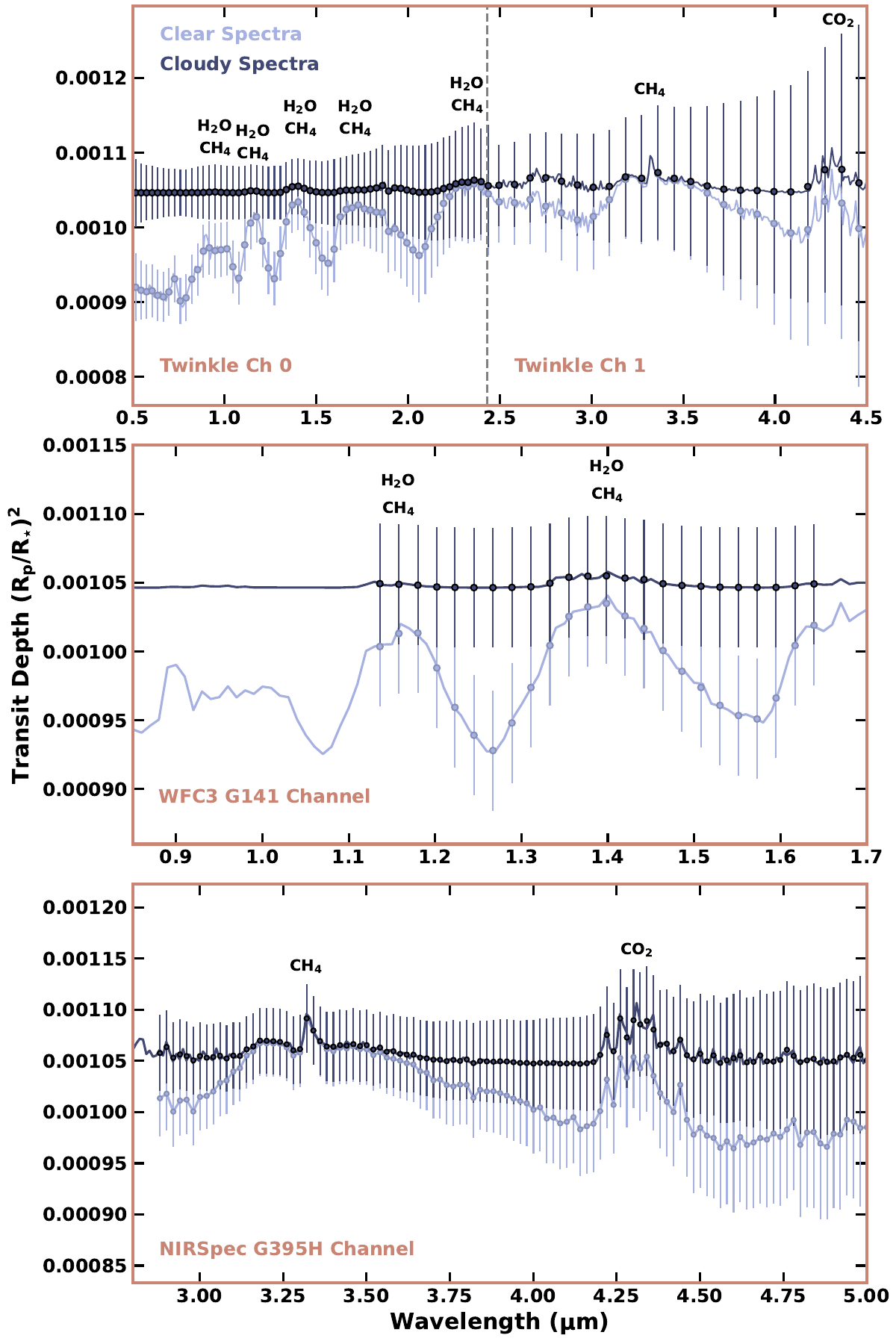}
    \caption{A synthetic atmospheric transmission spectrum of \targetc, with absorbing species including CH$_{4}$, CO, CO$_{2}$, H$_{2}$O, and NH$_{3}$ in chemical equilibrium at an atmospheric metallicity of $10\,\times$ Solar. Figure configuration is as per Figure~\ref{fig:tuarex_hip113103b}, with the \twinkle{} retrieval representing 10 visits.}
    \label{fig:tuarex_hip113103c}
\end{figure*}

\section{Investigating false positive scenarios} 
\label{sec:FP}
When identifying a new planetary system, it is important to carefully consider the possibilities of astrophysical and instrumental false positives.

When analysing beyond the \tess{} observations, \targetb{} and \targetc{} are detected with high significance on multiple instruments, yielding consistent transit depth and duration and thus sufficiently ruling out the scenario that the transit signals result from instrument false alarms of the \tess{} spacecraft.  

We use the following steps to rule out various astrophysical false positive scenarios. We can determine that either \targetb{} or \targetc{} are not eclipsing binaries around \target{} using radial velocity observations taken with CHIRON as outlined in Section~\ref{sec:spec}. There were no detections of significant radial velocity variations at the 20\,\ms{} level, ruling out stellar mass objects at the orbital period of the transit signals.  

We then follow \citet{2003_Seager} to use the transit shapes to constrain the probability that the transit signals were mimicked by a binary system blended with the \target{}. The maximum magnitude of an eclipsing binary that can produce a transit with similar shape can be estimated by:
\begin{equation}
    \Delta M \lesssim 2.5\,\text{log}_{10}\left(\frac{t_{12}^{2}}{t_{13}^{2}\delta}\right)^{2},
    \label{eqn:dmag}
\end{equation}
where $t_{12}$ represents the ingress duration, and $t_{13}$ represent the time between the first and the third contact of the transit.

We model the transit shape of both planets independently using the \tess{} and \cheops{} light curves without putting any priors on the stellar parameters. We found the 3--$\sigma$ $T_\mathrm{mag}$ upper limit of any background stars capable of producing these transit signals are 13  ($\Delta M<4.23$ mag) and 12  ($\Delta M<3.06$ mag) for \targetb{} and \targetc{}, respectively.

We rule out an hierarchical binary system associated with \target{}, satisfying the above criteria. Neither CHIRON nor the HARPS observations (Section~\ref{sec:spec}) detected secondary spectra lines, indicating no slow rotating, spectroscopic blended companions at $\Delta M<4$ \citep{2021AJ....161....2Z}.

For non-associated background binaries, we can use the \gaia{} DR3 catalogue to rule out stars brighter than our magnitude limit up to 1" away from \target. There are no stars within 20” of HIP 113103 based on Gaia DR3 catalogue.
We estimate the density of stars brighter than our magnitude limit within 1" of \target{} (which may be unresolved) by following the procedure described in \citet{2021AJ....161....2Z}. We found that the chance of finding a random star in the direction of \target{} with $\Delta M<4.2$ and $\Delta M<3$ are 3$\times10^{-5}$ and 1$\times10^{-5}$. 

Taken together, the combined observations that the system hosts multiple planets, the box-shaped transits, and the lack of additional stars in the spectra and background give us high confidence that \targetb{} and \targetc{} are genuine planets.

We also conducted a statistical validation on the \tess{} observations using the {\tt TRICERATOPS} package~\citep{2021_Giacalone}, the False Positive Probability (FPP) yielded 0.052 for \targetb{} and 0.026 for \targetc{}. The Nearby Star FPP (NFPP) for both planets is 0. The main contributor to the FPP is the scenario (STP) that a transiting planet with the same period is around an unresolved secondary star. We have high confidence that CHIRON and HARPS spectra can rule out secondary stars in the same system within $\upDelta$mag of 4, which is the magnitude limit to cause a transit given the transit shape constraints. The rest of false positive scenario have total FPP less than 1e-3, therefore we can confidently call both candidates planets.

\section{Results and Discussion}
\label{sec:discussion}

\subsection{Planet Properties}
\label{sec:planet_prop}

We statistically validate the planetary nature of the \target{} system, with the best fitted planetary parameters presented in Table~\ref{tab:planet_parameters}. \targetb{} has a radius of \Rp=\,\rpb \rearth, placing it in the upper bound of the radius gap, a small population of planets within the radii bounds 1.5\,\rearth$<$\,\Rp$\leq$\,2\,\rearth which may be the transition point from super-Earths to mini-Neptunes via photoevaporation-driven mass loss~\citep{2017_Fulton}. \targetc{} has a radius of \Rp=\,\rpc \rearth, and an equilibrium temperature of \teqc{}\,K, making it a warm mini-Neptune.

\begin{figure*}
    \centering
    \includegraphics[width=14cm]{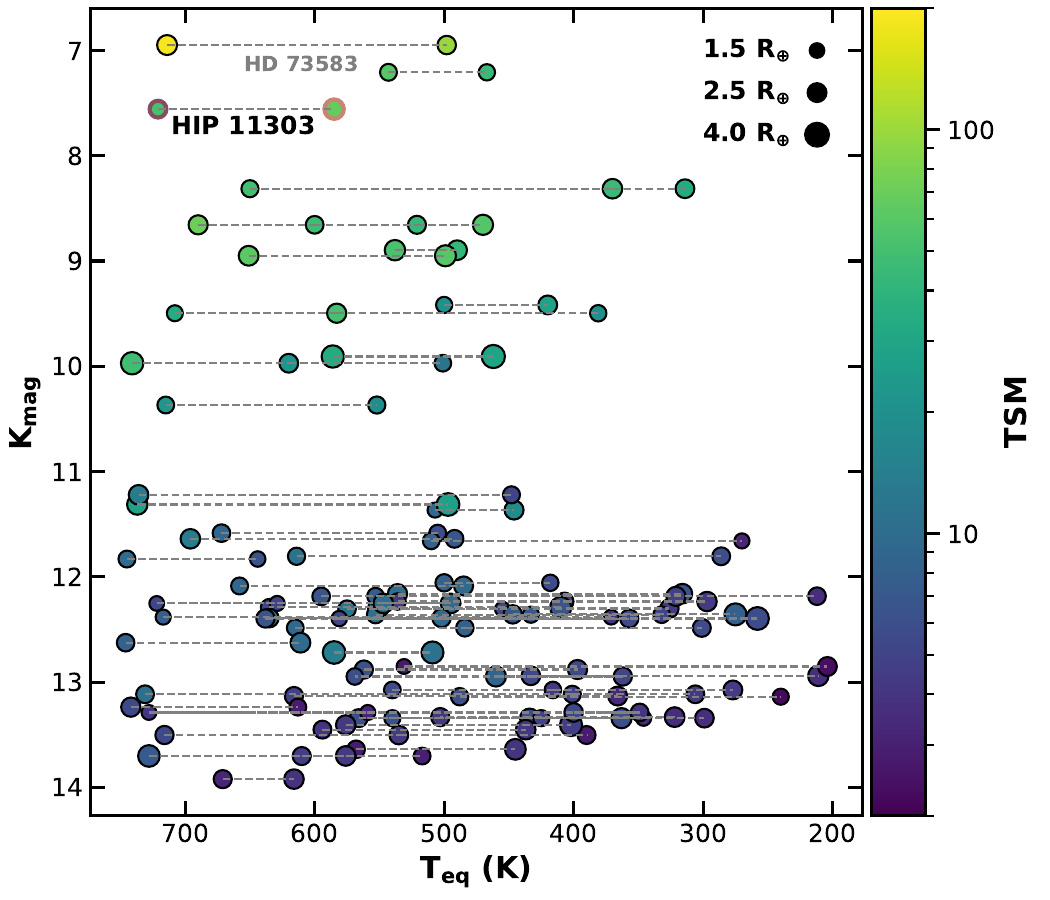}
    \caption{The \target{} system in the context of other multi-planet systems. Specifically, we show multi-planet systems hosting two or more warm Neptunes or super-Earths with equilibrium temperature \Teq{}$\leq750$ K, orbiting K stars as a function of their equilibrium temperature and K-band magnitude. The dashed lines connect each planet within the respective planetary system. The \target{} system orbits one of the brightest K-dwarf host stars, and are promising candidates for follow up atmospheric observations. The colour bar illustrating the transmission spectroscopy metric (TSM) for each planet places the \target{} system as second highest suitable for atmosphere analysis, behind the HD 73583 system~\citep{2022_Barragan}.}
    \label{fig:population}
\end{figure*}

We compare the \target{} system with other K stars hosting multi sub-Neptune planets with \Teq{} $\leq750$ K in (Figure~\ref{fig:population}), evaluating the transmission spectroscopy metric (TSM) for each target~\citep{2018_Kempton}. Due to the relative brightness of \target{} (V $\sim10$ mag) and high equilibrium temperatures, \targetb{} (TSM = 53) and \targetc{} (TSM = 68) are the second most suited system around a K star for atmosphere characterisation (only succeeded by the HD 73583 system~\citep{2022_Barragan}), and therefore invaluable targets to understand how multi sub-Neptune systems might evolve around K-stars. This stellar population is optimal for radial velocity due to its brightness in comparison to planets orbiting M-dwarfs~\citep{2018_Neil, 2023_Rojas-Ayala}. This aids the detection of smaller planets around a stellar population that shares similar characteristics to G-stars~\citep{2012_Howard}. Additionally from an atmosphere analysis perspective, K-stars have had repeated success at hosting planets with absorption at the He I 10830 \AA{} line, a tracer associated with atmosphere evaporation~\citep{2018_Nortmann, 2019_Allart, 2020_Guilluy, 2022_Fu}. Although the TSM values for \targetb{} and \targetc{} are below the J-band priority threshold of 90 (for targets within the 1.5 \rearth{} $\leq$ \Rp{} $\leq$ 10 \rearth{}), the derived radii of the planets combined with their close proximity to \target{}, and its Gyrochronological age make it a valuable system to explore through atmosphere analysis.

To estimate the mass of both planets to gauge the feasibility of future follow-up observations, we adopt the mass-radius relationship from \citet{2016_Wolfgang}. We estimate
$M_p=$ 5.9 $\pm$1.9 \mearth{} for \targetb{} and $M_p=$ 8.4 $\pm$1.9 \mearth{} for \targetc{}. These correspond to radial velocity semi-major amplitudes of 2.34 m s$^{-1}$ and 2.67 m s$^{-1}$, respectively. Additional analysis using the Echelle SPectrograph for Rocky Exoplanets and Stable Spectroscopic Observations ~\citep[ESPRESSO,][]{2021_Pepe} instrument on the Very Large Telescope (VLT) to understand this system in greater detail is currently underway.

\subsection{Transit Timing Variations} 
\begin{figure*}
    \centering
    \includegraphics[width=14cm]{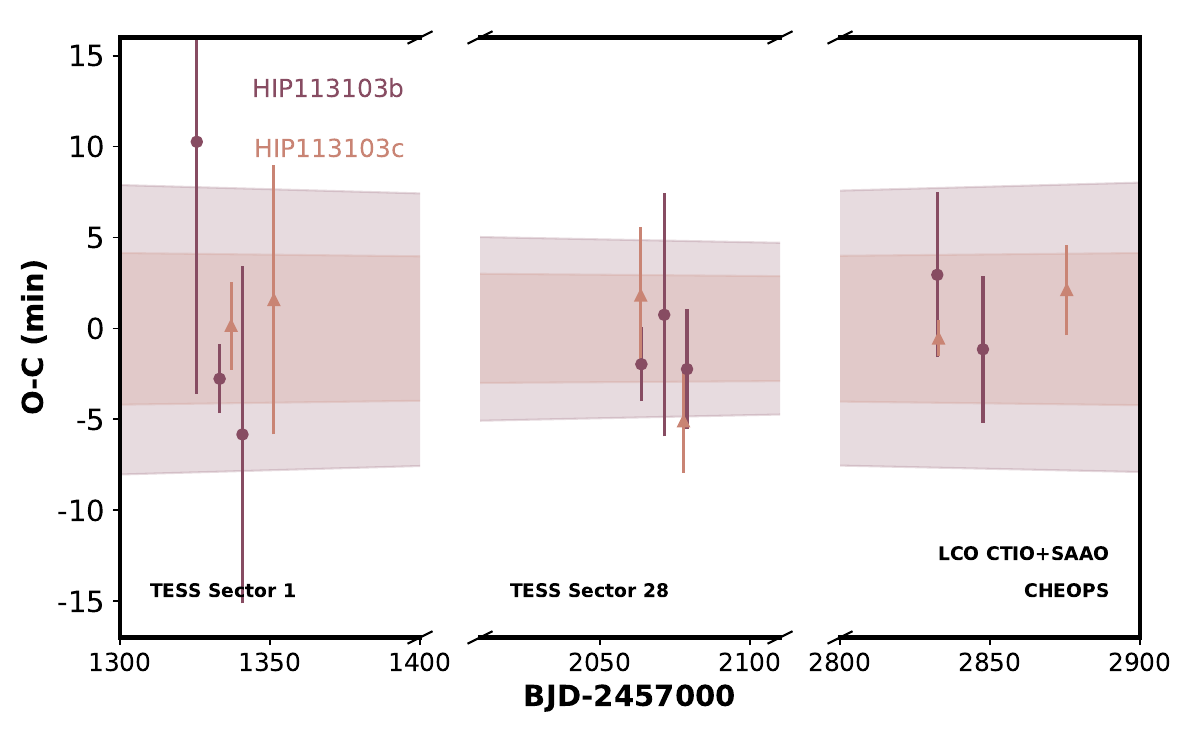}
    \caption{Deviations from linear transit times for individual transits of \targetb{} (circle markers) and \targetc{} (triangle markers) from all photometric observations, with each facility identified via labels along the time axis. Using our global fitted \To{} values to serve as the linear baseline for \targetb{} and \targetc{}, we show that over a four year period, the transit midpoint for each target does not vary beyond $\sim$10 minutes. shaded regions denote the $1\sigma$ propagated transit timing uncertainties for each respective planet. Note that the individual transit times are fully consistent with the linear transit ephemeris.}
    \label{fig:ttv}
\end{figure*}

We search for transit timing variations (TTVs), indicative of interactions between the two planets, or the presence of additional companions. To derive accurate transit times for each event, we perform an additional global model of the system, as per Section~\ref{sec:global_model}, where the transit epoch of each transit event is a free variable, and the period is held fixed. The resulting transit times are displayed in Figure~\ref{fig:ttv}. We find no evidence for deviations from a linear ephemeris propagation larger than 4.5 (resp. 2.5) minutes (1-$\sigma$ scatter) for \targetb{} (resp. c). The mean timing uncertainty per transit is 5.7 (resp. 3.3) minutes, consistent with the measured scatter. We estimated the expected TTV amplitude for the system using \texttt{TTVFaster} \citep{2016ApJ...818..177A}. Given the pair of planets has a period ratio within 10\% of the 2:1 mean motion resonance, we can estimate that if the planets are on modest eccentric orbits or are relatively massive, they are highly likely to exhibit TTVs with amplitudes detectable by our observations. In detail, if both planets are of Neptune mass ($\sim$$17\,M_{\oplus}$), and we assume relatively low eccentricities follow a Rayleigh distribution with $\sigma{_e}=0.06\,(\text{consequent mean eccentricity,}\,\bar{e} = 0.075)$, the median TTV scatter for \targetb{} (resp. \targetc{}) should be on the 10\,(resp. 15) minutes timescale~\citep{2008_Juric}. If we adopt masses for both planets as per the mass-radius relationship from \citet{2016_Wolfgang}, \texttt{TTVFaster} estimates that the eccentricity of the system is most likely lower than 0.2 given the non-detection of a significant TTV signal. This puts tighter constraints on the system eccentricity compared to our global modeling. 
This eccentricity upper limit is derived by comparing the scatter in the transit times from \texttt{TTVFaster} simulations and the observed data. We first compute the 3 sigma scatter of the transit times from the dataset presented in Figure 12. We then compute the corresponding eccentricities of the systems that would produce larger transit time deviations in 95\% of the simulations.

\subsection{Prospect for Atmospheric Follow-up} 
Systems hosting transiting small planets are optimal for understanding the radius evolution and mass-loss processes that sculpt the close-in sub-Neptune population \citep[e.g.][]{2020MNRAS.491.5287O}. Having planets that formed from the same protoplanetary disk and experienced the same host star XUV evolution, allows us to test photoevaporation processes by isolating the effects of insulation on mass-loss. We tested for the future prospect of atmospheric transmission spectroscopic observations for the planets in the \target{} system via a set of current and upcoming space-based facilities, including the \textit{Twinkle Space Telescope} (\textit{Twinkle})~\citep{2022_twinkle}, \hst{}~\citep{2008_Kimble, 2016_Tsiaras}, and \jwst{}~\citep{2018_Bean, 2022_Jakobsen}. \twinkle{} is a visible to infrared ($0.5\,\upmu$m $- 4.5\,\upmu$m) 0.45\,m space telescope set to begin scientific operations in 2025 at a 700\,km geocentric Sun-synchronous orbit.

The simulated transmission \twinkle{} spectra is generated for both channels ($0.5\,\upmu$m $\leq$ Ch0 $\leq2.43\,\upmu$m, $2.43\,\upmu$m $\leq$ Ch1 $\leq4.5\,\upmu$m) using the radiometric tool available on the mission's database Stardrive\footnote{\href{https://stardrive.twinkle-spacemission.co.uk/}{Stardrive Database}}, while the \hst{} (WFC3 NIR G141 grism: $1.075\,\upmu$m $-1.7\,\upmu$m) and \jwst{} (NIRSpec G395H grism: $2.87\,\upmu$m $-5.14\,\upmu$m) transmission spectra are generated using the publicly available noise simulator, \texttt{PandExo}~\citep{2017_Batalha}. The synthetic transmission spectra are processed through the retrieval framework \texttt{TauREx 3.0}~\citep{2021_Al-Refaie}, which generates an atmosphere divided into 100 evenly spaced layers across a log grid varying from $10^{-4}$ to $10^{6}$ Pa. The trace gases in our models were assumed to be in chemical equilibrium and the abundances were calculated using \texttt{FastChem}~\citep{2018_Stock}. We keep the C/O ratio fixed at 0.54 \citep[an oxygen rich atmosphere at solar abundances as discussed in][]{2012_Madhusudhan} and assume a metallicity of $100\,\times$ Solar for \targetb{} and $10\,\times$ Solar for \targetc{}, aligning with previous studies that relate metallicity and low-mass planets as being inversely proportional~\citep[e.g. ][]{2013_Fortney, 2014_Kreidberg, 2015_Charnay}. The trace gases inserted into this atmosphere included the molecular opacities of CH$_{4}$~\citep{hill_xsec, exomol_ch4}, CO~\citep{2015_Li}, CO$_{2}$~\citep{2010_Rothman}, H$_{2}$O~\citep{2018_Polyansky}, and NH$_{3}$~\citep{Yurchenko_2011_NH3}, with all opacities being obtained through the ExoMol~\citep{2016_Tennyson} and HITRAN databases~\citep{2016_Gordon}. In addition, we also implement Rayleigh scattering for all inserted molecules~\citep{cox_allen_rayleigh}, provide a collision induced absorption from H$_{2}$-H$_{2}$~\citep{2011_Abel, 2018_Fletcher} and H$_{2}$-He interactions~\citep{2012_Abel}. For each instrument, we modelled both a clear atmosphere scenario (i.e. $P=10^{6}$ Pa) and a uniform opaque deck scenario (i.e. grey clouds) at $P=10^{1}$ Pa. 

We present our simulated spectra of \targetb{} in Figure~\ref{fig:tuarex_hip113103b} and \targetc{} in Figure~\ref{fig:tuarex_hip113103c}. Should \targetb{} and \targetc{} reflect a similar composition as our simulated spectra, we can recover a transmission spectra from 10 orbits using \twinkle{} (top panels)~\citep{2022_twinkle}. Likewise, Figures~\ref{fig:tuarex_hip113103b} and~\ref{fig:tuarex_hip113103c} demonstrate the precision we can expect to achieve from one orbit observation using the infrared WFC3 G141 grism (middle panels) on \hst{}~\citep{2008_Kimble, 2016_Tsiaras} and NIRSpec G395H grism (bottom panels) on \jwst~\citep{2018_Bean, 2022_Jakobsen}. We can successfully retrieve molecular species for \targetc{} with each instrument given a clear atmosphere (with H$_{2}$O and CO$_{2}$ displaying the strongest absorption). However, in the event of clouds, we would struggle to detect any signal for \targetc{} using \twinkle{} and \hst{} instruments. Distinguishing between a clear or cloudy atmosphere on all three instruments is challenging for \targetb{}. An atmosphere that is in chemical disequilibrium for both targets could result in stronger absorption but would be dependent on various unknown physical parameters. Alternatively, the evolutionary path of \targetb{} could be the product of a migrated Water World instead of photoevaporation of a mini-Neptune~\citep[e.g.][]{2022_Luque}, however this scenario can only be explored in more detail after planet density measurements have been calculated. Our density estimation for \targetb{} of \rhob{} \rhoe{} is indicative of a rocky planet, but is inferred from a mass-radius relationship and may not reflect the true bulk density of the planet. Accurate mass measurements of \targetb{} and \targetc{} are required to confirm their densities.

\section{Conclusion}
\label{sec:conclusion}
In this paper, we confirm the existence of two sub-Neptunes, \targetb{} and \targetc{}, within $\sim$10$\%$ of 2:1 resonance around the bright K3V star \target{}. First identified with \tess{}, this system is revisited using both ground based transit observations (observed with the photometric LCO network and on the CHIRON spectrograph), as well as a space-based photometric observations of both targets within a $\sim$17.5 hour visit using \cheops{}. Follow up TTV analysis does not reveal any additional outer companions. Our planetary parameters revealed a radius of~\Rp=\,\rpb \rearth for \targetb{} and \Rp=\,\rpc \rearth for \targetc{}, confirming both targets reside within the mini-Neptune sub-class. For \targetb{}, the combination of its close proximity to \target{} and its planetary radius means it resides within the radius gap, which if confirmed via mass follow up, would add an additional target a sparse sub-class of planets which are hypothesised to bridge the formation transition between super-Earths and mini-Neptunes. If \targetb{} is the subject of atmosphere evaporation due to its close proximity to \target{}, our generated retrieval plots (using the \twinkle{}, \hst{}, and \jwst{} telescopes) suggest it would be a struggle to distinguish an evolutionary gap via metallicity disparity for all three telescopes (assuming chemical equilibrium), even if there is a clear atmosphere. Ultimately, this system provides two key targets capable of atmospheric analysis within the population of mini-Neptune multi-planet systems orbiting K-stars.

\section*{Acknowledgements}
We would like to acknowledge and pay respect to Australia’s Aboriginal and Torres Strait Islander peoples, who are the traditional custodians of the lands, the waterways and the skies all across Australia. We thank Australia’s Aboriginal and Torres Strait Islander peoples for sharing and caring for the land on which we are able to learn. In particular, we pay our deepest respects to all Elders, ancestors and descendants of the Giabal, Jarowair, and Yuggera nations, upon which the \textsc{Minerva}-Australis facility is situated, and analysis for this paper was undertaken. We would also like to acknowledge and pay our deepest respects to the  Indigenous Elders, ancestors and descendants who are the traditional custodians of the land upon which the CTIO, SAAO, and ESO 3.6\,m are situated. This includes (but may not be exclusive to) the Diaguita and Khoisan nations. 
GZ thanks the support of the ARC DECRA program DE210101893. CXH thanks the support of the ARC DECRA program DE200101840.
GZ, SQ thank the support of the \tess{} Guest Investigator Program G03007. 
CH thanks the support of the ARC DECRA program DE200101840.
KAC acknowledges support from the \tess{} mission via subaward s3449 from MIT. This research has used data from the CTIO/SMARTS 1.5m telescope, which is
operated as part of the SMARTS Consortium by \href{http://secure-web.cisco.com/1TL5nionOJJUGi7T0X_YvX7RLRwbVQl20QG7s4LKeK1vpFY8M3UHYMuONVvV2D2hxli_pMi4YkHdTYel4ogZ3sJWN4axM8-5IsyCIPeIj7BfVIBOvp9a8iRKv2IM-wTBpjGA3xxZcH5lT4FNKBIoEstyJEEyUYzEKbDQyL4T1LQSiukl5eTarVlkS9YJbHf_HrjiuXV1gM1uXr7gdIdCbZg4CfJa_N8Qw38oz0KhpJ74RZ0rIcyg3XKCc6-HCDjlBrMtX3cpMKa1Kcya1SxY0UxXY0WkwM0zGeXYUYfbkp1Ce6jIBY8Evcz-YcyODRE4QWMlPqSDV66bKv5F1R3-RrkcH91Y7INyFOP6qJfGJKLRFJT-KNphpqmNc4Pf7zLVOIBjCEKsANmt1XTtzQN5AIPwKf-F1qd4b6KCZrqjHZIA/http\%3A\%2F\%2Fwww.recons.org}{RECONS}.
This work makes use of data from the European Space Agency (ESA) mission \cheops{} via the \cheops{} Guest Observers Program AO-3-10.
This work has made use of data from the European Space Agency (ESA) mission
\href{https://www.cosmos.esa.int/gaia}{\it Gaia}, processed by the {\it Gaia}
Data Processing and Analysis Consortium \href{https://www.cosmos.esa.int/web/gaia/dpac/consortium}{(DPAC)}. Funding for the DPAC
has been provided by national institutions, in particular the institutions
participating in the {\it Gaia} Multilateral Agreement.
This work makes use of observations from the LCOGT network. Part of the LCOGT telescope time was granted by NOIRLab through the Mid-Scale Innovations Program (MSIP). MSIP is funded by NSF. This work makes use of data from the \textsc{Minerva}-Australis facility. \textsc{Minerva}-Australis is supported by Australian Research Council LIEF Grant LE160100001, Discovery Grants DP180100972 and DP220100365, Mount Cuba Astronomical Foundation, and institutional partners University of Southern Queensland, UNSW Sydney, MIT, Nanjing University, George Mason University, University of Louisville, University of California Riverside, University of Florida, and The University of Texas at Austin.
This research has made use of the NASA Exoplanet
Archive, which is operated by the California Institute of Technology, under contract with the National Aeronautics and Space Administration under the Exoplanet Exploration Program. This work has been carried out within the framework of the NCCR PlanetS supported by the Swiss National Science Foundation under grants 51NF40$_{182901}$ and 51NF40$_{205606}$. Funding for the \tess{} mission is provided by NASA's Science Mission directorate. We acknowledge the use of public \tess{} Alert data from pipelines at the \tess{} Science Office and at the \tess{} Science Processing Operations Center. This research has made use of the Exoplanet Follow-up Observation Program (ExoFOP; DOI: 10.26134/ExoFOP5) website, which is operated by the California Institute of Technology, under contract with the National Aeronautics and Space Administration under the Exoplanet Exploration Program. This paper includes data collected by the \tess{} mission, which are publicly available from the Mikulski Archive for Space Telescopes (MAST).
Resources supporting this work were provided by the NASA High-End Computing (HEC) Program through the NASA Advanced Supercomputing (NAS) Division at Ames Research Center for the production of the SPOC data products.

\section*{Data Availability}
The data underlying this article will be shared on reasonable request to the corresponding author.



\bibliographystyle{mnras}
\bibliography{ref} 




\bsp	
\label{lastpage}
\end{document}